\documentclass[10pt]{article}
\usepackage[verbose, a4paper, hmargin=2.5cm, vmargin=2.5cm]{geometry}

\usepackage{paratype}
\usepackage{times}
\usepackage{amsmath}
\usepackage{amsfonts}
\usepackage{amssymb}
\usepackage{esint}

\usepackage{graphicx}
\usepackage[export]{adjustbox}
\usepackage{mdframed}
\usepackage{caption}
\usepackage{booktabs,array,multirow}
\usepackage{adjustbox}
\usepackage[hyphens]{url}

\usepackage{hyperref}
\hypersetup{colorlinks=true, linkcolor=blue, filecolor=magenta, urlcolor=cyan,}
\graphicspath{ {./images/} }

\newcommand{\customfootnote}[1]{
  \let\thefootnote\relax\footnotetext{#1}
}

\title{Deep Learning Models for Colloidal Nanocrystal Synthesis}
\author{
    Kai Gu\(^1\textsuperscript{\#}\), Yingping Liang\(^2\textsuperscript{\#}\), Jiaming Su\(^1\),
    Peihan Sun\(^1\), Jia Peng\(^1\), Naihua Miao\(^3\),\\
    Zhimei Sun\(^3\), Ying Fu\(^2\textsuperscript{*}\), Haizheng Zhong\(^1\textsuperscript{*}\),
    Jun Zhang\(^4\)
}
\date{}

\begin{document}

\maketitle

\begin{center}
\(^1\) MIIT Key Laboratory for Low-Dimensional Quantum Structure and Devices, School of Materials Sciences \& Engineering, Beijing Institute of Technology, Beijing 100081, China.\\[6pt]

\(^2\) School of Computer Science and Technology, Beijing Institute of Technology, Beijing 100081, China.\\[6pt]

\(^3\) School of Materials Science and Engineering, Beihang University, Beijing 100191, China.\\[6pt]

\(^4\) State Key Laboratory of CNS/ATM \& MIIT Key Laboratory of Complex-field Intelligent Sensing, Beijing Institute of Technology, Beijing 100081, China.\\[12pt]

E-mail: \texttt{hzzhong@bit.edu.cn}; \texttt{fuying@bit.edu.cn}\\[6pt]
\end{center}

\begin{flushleft}
\(\#\) These authors contributed equally: Kai Gu, Yingping Liang.
\end{flushleft}

\section*{Abstract}

Colloidal synthesis of nanocrystals usually includes complex chemical reactions and multi-step crystallization processes. Despite the great success in the past 30 years, it remains challenging to clarify the correlations between synthetic parameters of chemical reaction and physical properties of nanocrystals. Here, we developed a deep learning-based nanocrystal synthesis model that correlates synthetic parameters with the final size and shape of target nanocrystals, using a dataset of 3500 recipes covering 348 distinct nanocrystal compositions. The size and shape labels were obtained from transmission electron microscope images using a segmentation model trained with a semi-supervised algorithm on a dataset comprising 1.2 million nanocrystals. By applying the reaction intermediate-based data augmentation method and elaborated descriptors, the synthesis model was able to predict nanocrystal’s size with a mean absolute error of 1.39 nm, while reaching an 89\% average accuracy for shape classification. The synthesis model shows knowledge transfer capabilities across different nanocrystals with inputs of new recipes. With that, the influence of chemicals on the final size of nanocrystals was further evaluated, revealing the importance order of nanocrystal composition, precursor or ligand, and solvent. Overall, the deep learning-based nanocrystal synthesis model offers a powerful tool to expedite the development of high-quality nanocrystals.

\section*{Main}

Colloidal nanocrystals are a representative class of nanomaterials, widely applied in many cutting-edge technologies \({}^{1}\) such as optoelectronics \({}^{2}\) , energy catalysis \({}^{3,4}\) , and biomedicine \({}^{5}\) due to their pronounced size-dependent effects \({}^{6 - 8}\) . Over the past 30 years, advances in colloidal chemistry have led to the development of numerous synthetic methods and recipes for fabricating nanocrystals with controllable size and shape \({}^{9 - {13}}\) . Some nanocrystals, such as CdSe and InP quantum dots, have even achieved scale-up production \({}^{{14} - {17}}\) . A typical nanocrystal synthesis involves three main crystallization processes: conversion of precursors to monomers (pre-nucleation), aggregation of monomers into nuclei (nucleation), and growth of nanocrystals through monomer diffusion \({}^{{18},{19}}\) . The final size and shape of nanocrystals are strongly correlated with these \({\text{ processes }}^{{20},{21}}\) . However, the intrinsic relationship between chemical reactions and crystallization processes remains poorly understood. It is still challenging to quantitatively describe how synthetic parameters (temperature, reactant ratios, and ligand types, etc.) affect the final size and shape of nanocrystals.

Machine learning provides a powerful approach for establishing mathematical relationships among complex variables, showing potential in material design and property prediction \({}^{{22} - {28}}\) . In the context of nanocrystal synthesis \({}^{{29},{30}}\) , previous studies have reported the use of machine learning to predict the size and absorption spectra of nanocrystals \({}^{{31} - {33}}\) . Due to the lack of comprehensive datasets and generalizable descriptors, these machine learning models have limited knowledge transfer capabilities to predict the properties of target nanocrystals. In this work, a dataset comprising 3500 synthesis recipes involving 348 types of nanocrystals and a dataset containing 12000 transmission electron microscope (TEM) images (1.2 million nanocrystals) were constructed by extracting data from literature and experiments. Utilizing the nanocrystal sizes obtained from the segmentation model and shapes as labels, and the synthetic parameters and chemical structures extracted from the recipe dataset as input, a deep learning-based nanocrystal synthesis model was developed to achieve size and shape predictions from the synthesis recipes.

Fig. 1 provides a schematic overview of the process for developing the nanocrystal segmentation and synthesis models. The segmentation model for nanocrystals was trained using a segmentation network combined with a semi-supervised learning strategy, enabling the precise size measurement and shape clustering of nanocrystals.

The semi-supervised learning strategy significantly enhanced the accuracy of nanocrystal localization, achieving in an average segmentation precision of 82.5\%. The segmentation model provides size labels for constructing the synthesis model. The recipe dataset includes various synthetic methods for nanocrystals, where reaction conditions (such as temperature, time, and chemical concentration) serve as condition descriptors, and three-dimensional (3D) chemical structures (including the structures of precursors, ligands, and solvents) act as chemical descriptors. A tenfold augmentation of input data was achieved by applying a reaction intermediate-based data augmentation method. This data augmentation, combined with deep learning, enhances the accuracy of the synthesis models in predicting the size and shape of nanocrystals. The mean absolute error (MAE) for size prediction is \({1.39}\mathrm{\;{nm}}\) , and the average accuracy for shape classification is \({89}\%\) . In generalizability tests, this synthesis model also shows knowledge transfer capabilities across different nanocrystals. By analyzing the self-attention weights of the model, we found that the chemicals in nanocrystal synthesis follow an order of importance: nanocrystal composition, precursor or ligand, and solvent.

\begin{figure}[htbp]
    \centering
    \includegraphics[max width=1.0\textwidth]{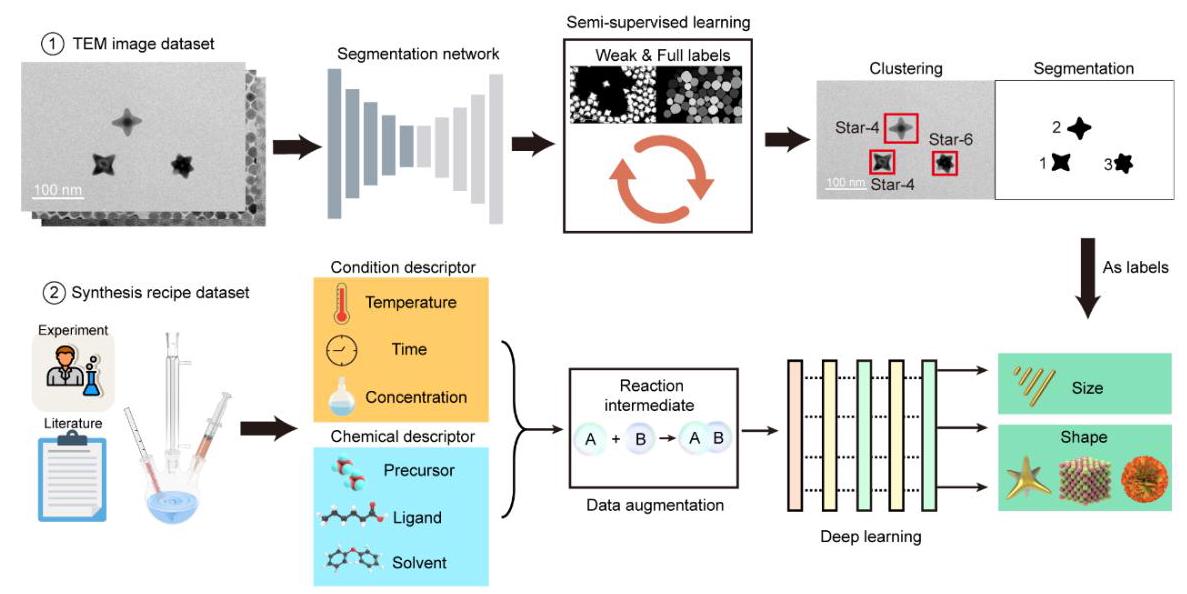}
    \caption*{Fig. 1 | Schematic diagram of the segmentation and synthesis models. Upper panel: The segmentation model for nanocrystals was trained using a segmentation network combined with a semi-supervised learning strategy on the TEM image dataset, enabling instance segmentation and shape clustering of nanocrystals. The sizes obtained from the segmentation model were utilized as labels for training the synthesis model. Bottom panel: The synthesis model of nanocrystals was trained using deep learning algorithms. Condition descriptors and chemical descriptors were extracted from the synthesis recipe dataset and used as input features to predict nanocrystal size and shape with the reaction intermediate-based data augmentation method.}
    \label{fig:segmentation_synthesis}
\end{figure}

A precise size determination of nanocrystals is essential for constructing the synthesis model. Compared with manual measurement, deep learning-based segmentation of nanocrystals from TEM image provides an accurate method for size determination \({}^{{34} - {37}}\) . To address the challenges of high particle density and low pixel resolution of nanocrystals in TEM images, we proposed a semi-supervised learning segmentation model, which can efficiently and accurately obtain the size and shape of nanocrystals. Fig. 2a schematically illustrates the semi-supervised learning process. It consists of two distinct training branches: one for labeled images and another for unlabeled images with weak label generation. In the labeled image branch, fully labeled TEM images were processed through an encoder and a seed decoder to generate seed maps that indicate the probability of finding a nanocrystal, represented by different colors. These seed maps were further refined by an instance decoder to produce instance maps, and the loss was calculated by comparing these instance maps to the full labels. For the unlabeled image branch, weak labels were generated by a pre-trained object detection network and morphological operations, which were categorized into foreground, background, and uncertain regions, as shown in Extended Data Fig. 1. The binary cross-entropy loss was calculated using the foreground and background regions, as detailed in Methods. The training dataset comprises both labeled images with full labels and unlabeled images, generating a total of 900,000 nanocrystals with weak labels, which is approximately twenty times the number of full labels (Supplementary Table 1). By incorporating the unlabeled images, the segmentation model improved the average precision (AP), with the value of \({\mathrm{{AP}}}_{50}\) increasing from \({68.5}\%\) to \({80.1}\%\) (definition of \({\mathrm{{AP}}}_{50}\) , see Methods), as shown in Extended Data Fig. 2. By incorporating additional full labels from literature, the average precision was increased to 82.5\%. This enhancement was also confirmed by other metrics including \({\mathrm{{AP}}}_{75}\) and \({\mathrm{{AP}}}_{90}\) (see Methods). Extended Data Fig. 3 demonstrates that the addition of unlabeled images improved the localization of nanocrystals, thereby enhancing segmentation accuracy, particularly for smaller-sized nanocrystals. By further applying a confidence filter, the segmentation model can remove low-confidence areas and automatic segmentation of large numbers of nanocrystals effectively ensuring precise description of the size and shape.

Fig. 2b and Extended Data Fig. 4 present TEM images before and after nanocrystal segmentation, showing accurate segmentation of nanocrystals with different shapes, sizes, and aggregation states. Based on the segmented images, the size, size distribution, and counts of nanocrystals can be calculated (refer to Supplementary Fig. 1 and Methods for size calculation). The segmented nanocrystals were further analyzed using five shape descriptors including circularity, solidity, convexity, eccentricity and aspect ratio (Supplementary Fig. 2), providing a tool to analyze the shapes of nanocrystals. For instance, the shapes of 130,000 PbSe nanocrystals synthesized with different recipes were clustered into seven groups (C1 to C7), as shown in Fig. 2c. Notably, C1 is positioned near all other clusters, suggesting its intermediate role in shape evolution. Furthermore, the 130,000 PbSe nanocrystals were categorized according to reaction temperatures of \({140}{}^{ \circ  }\mathrm{C},{160}{}^{ \circ  }\mathrm{C},{180}{}^{ \circ  }\mathrm{C}\) , and \({240}{}^{ \circ  }\mathrm{C}\) (Supplementary Fig. 3). The proportion distribution of shapes varies with reaction temperature. Fig. 2d summarizes the proportions of the seven nanocrystal shapes synthesized at different reaction temperatures. Star-shaped nanocrystals are only observed at \({180}{}^{ \circ  }\mathrm{C}\) , while the square shape predominates at \({240}^{ \circ  }\mathrm{C}\) .

\begin{figure}[htbp]
    \centering
    \includegraphics[max width=1.0\textwidth]{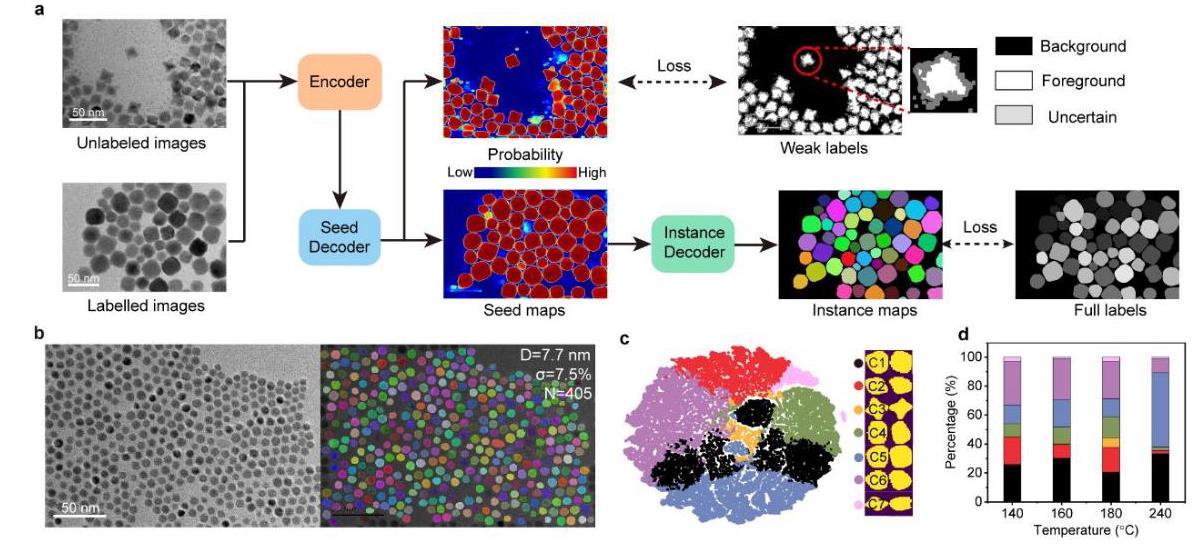}
    \caption*{Fig. 2 | Segmentation and shape analysis of nanocrystals from TEM images. \textbf{a}. Schematic of the semi-supervised learning process for nanocrystal segmentation. \textbf{b}. TEM images before and after nanocrystal segmentation. \textbf{c}. Clustering and visualization of 130,000 PbSe nanocrystals based on shape descriptors, where different colors represent distinct shapes: black (C1, irregular), red (C2), orange (C3, star), green (C4), blue (C5, square), purple (C6), and pink (C7). \textbf{d}. Proportions of the seven nanocrystal shapes synthesized at different reaction temperatures, with colors corresponding to the shape clusters.}
\end{figure}

Descriptors extracted from the synthesis recipes serve as inputs for constructing the synthesis model. Fig. 3a presents a schematic of the data processing workflow of input descriptors, including condition descriptors and chemical descriptors derived from the recipe dataset. The condition descriptor, which includes five reaction parameters, provides a simplified representation of different synthesis methods (Supplementary Fig. 4). To obtain chemical descriptors, the chemical names from the recipes were converted into 3D structures using density functional theory (DFT) calculations. These 3D structures were then incorporated into a graph neural network (GNN) to generate chemical descriptors, as illustrated in Extended Data Fig. 5a (see Methods for further details). Given that larger datasets improve the performance of deep learning models, we developed a reaction intermediate-based data augmentation method to improve the model's generalization. This approach involves using DFT calculations to derive chemical descriptors for the reaction intermediates of any two chemicals within a recipe, and subsequently updating the original recipe, resulting in a tenfold data augmentation (Extended Data Fig. 5b). As shown in Extended Data Fig. 6, the recipe dataset includes 3500 synthesis recipes covering 348 nanocrystals compositions, an order of magnitude larger than previous reports \({}^{{31} - {33}}\) . Among them,562 synthesis recipes for elemental substances, 1961 for binary compounds, 745 for ternary compounds and 225 for quaternary compounds. Furthermore, the recipe dataset includes 382 different precursors and 137 different ligands. Extended Data Fig. 6c shows the elemental frequency distribution of nanocrystal compositions in the recipe dataset, covering a wide range of the periodic table.

Using this comprehensive dataset, we proposed a transformer-based synthesis model to predict the size and shape of nanocrystals. As shown in Extended Data Table 1, the transformer-based model with the data augmentation achieved an MAE of 1.39 \(\mathrm{{nm}}\) and \({\mathrm{R}}^{2}\) of 71.9\% for size prediction (Fig. 3b), and the average accuracy of \({89}\%\) for shape classification (Extended Data Fig. 7 and Supplementary Table 2). As shown in Fig. 3c, the input descriptors of synthesis model are naturally divided into ten distinct clusters in two-dimensional space using the t-distributed stochastic neighborhood embedding method, where the colors represent the shape labels. Notably, the resulting cluster distribution are similar with the distribution of shape labels, confirming the effective description of synthesis recipes. The synthesis model has the capability to optimize recipes for fabricating nanocrystals with targeted size and shape. Taking the synthesis of PbSe nanocrystals for example, the size of PbSe nanocrystals can be finely tuned from 8 to \({18}\mathrm{\;{nm}}\) by varying the molar amount of oleic acid (OA) and oleylamine (OLA), as shown in Fig. 3d. Meanwhile, their shapes can be varied among sphere, rhombus, and cube (Fig. 3e). In addition, we also developed the reverse design capability of synthesis model to predict the molar amounts of reactants for desired size, achieving an MAE of 2.41 mmol (Supplementary Table 3).

\begin{figure}[htbp]
    \centering
    \includegraphics[max width=1.0\textwidth]{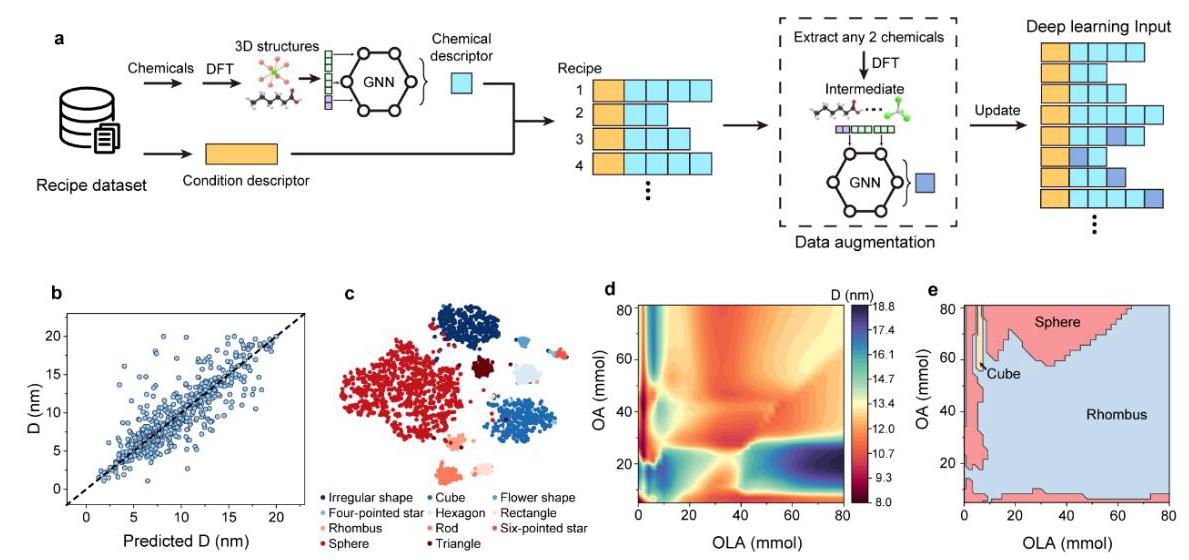}
    \caption*{Fig. 3 | Development and evaluation of the nanocrystal synthesis model. a, Data processing procedure for input features to the synthesis model. b, Parity plot of size prediction using the transformer-based model. c, Visualization of input descriptors to the synthesis model using t-distributed stochastic neighbor embedding method. d, Size 2 and (e) shape evolution of PbSe nanocrystals with different molar amount of OLA and OA.}
\end{figure}

Furthermore, we evaluated the transfer capability of synthesis model across different nanocrystal recipes through generalizability tests. All recipes for specific nanocrystal compositions were selected as a validation set to evaluate the model's predictive potential with new recipes as input (see Methods for details). For example, all recipes of PbSe nanocrystals were selected as a test set, and the prediction performance of the synthesis model is shown in Extended Data Table 2. Despite the exclusion of PbSe recipes, the model successfully predicted the size of PbSe nanocrystals with given recipes, achieving an MAE of \({2.65}\mathrm{\;{nm}}\) . The transfer capability was also confirmed by generalizability tests on datasets excluding \({\mathrm{{Ag}}}_{2}\mathrm{\;S},\mathrm{{Ni}}\) or \({\mathrm{{Cs}}}_{2}{\mathrm{{NaBiCl}}}_{6}\) nanocrystals. We also explored the transfer capability of synthesis model using recipes published from the year 2023 to 2024 as a testing set. As summarized in Supplementary Table 4, the true and model-predicted sizes of the reported \({\mathrm{{CaF}}}_{2}\) 04 nanocrystals are 9.2 and \({9.82}\mathrm{\;{nm}}\) , respectively \({}^{38}\) . These results suggest that the transformer-based synthesis model effectively transfers knowledge across different 206 nanocrystal recipes.

Chemical reactions and crystallization processes are two important topics in nanocrystal synthesis, while the theoretical models mainly focused on crystallization. We can quantify the effect of condition descriptor and molar amount of the chemical on the size prediction, by Shapley value analyses (Extended Data Fig. 8) \({}^{39}\) . To understand the role of chemicals in reactions, we further explored the interpretability of synthesis model via the self-attention mechanism in the transformer algorithm. The attention weights can quantify the interactions between different chemicals. For instance, in the synthesis of PbSe nanocrystals using lead oxide (PbO), selenium powder (Se), OA, OLA, octadecene (ODE), and tri-n-octylphosphine (TOP), the attention weights were analyzed across different layers of the transformer model. As shown in Fig.4a, the attention weights between Se and OA (block i), OA and OLA (block ii) exhibited the high values in the first layer. In the second layer, PbO and OA (block iii), Se and TOP (block iv) exhibited high attention weight values. OLA showed higher attention weights with other chemicals compared to others in the third layer, while TOP exhibited higher weights in the fourth layer. These findings align with known reaction mechanisms, where \(\mathrm{{PbO}}\) and \(\mathrm{{OA}}\) react to form reactive precursors, and Se interacts with TOP or OA to initiate key reactions in nanocrystal formation \({}^{{40},{41}}\) .

A learnable vector class token (CLS) within the transformer structure was introduced as a descriptor for chemical reactions (Supplementary Fig. 5). The attention weights between CLS and different chemicals are used to evaluate the importance of each chemical in chemical reactions. Supplementary Fig. 6 shows the average attention weights across different transformer layers, while Fig. 4b shows the corresponding maximum average attention weights (see Methods for calculation details). Most chemicals are concentrated in the first and second transformer layers. Chemicals in the nanocrystal synthesis are usually categorized into nanocrystal compositions, precursors, solvents and ligands. Fig. 4c summarizes the categorization and distribution of chemicals in different layers. The nanocrystal compositions are primarily located at the first layer, while the precursors and ligands concentrate in the second and third layers.

Solvents are mainly present in the fourth and fifth layers. The similarity of solvent and ligand distributions can be explained to their interchangeable roles in chemical reactions. Overall, the chemicals follow an order of importance: nanocrystal composition, precursor or ligand, solvent in determining the final size.

\begin{figure}[htbp]
    \centering
    \includegraphics[max width=1.0\textwidth]{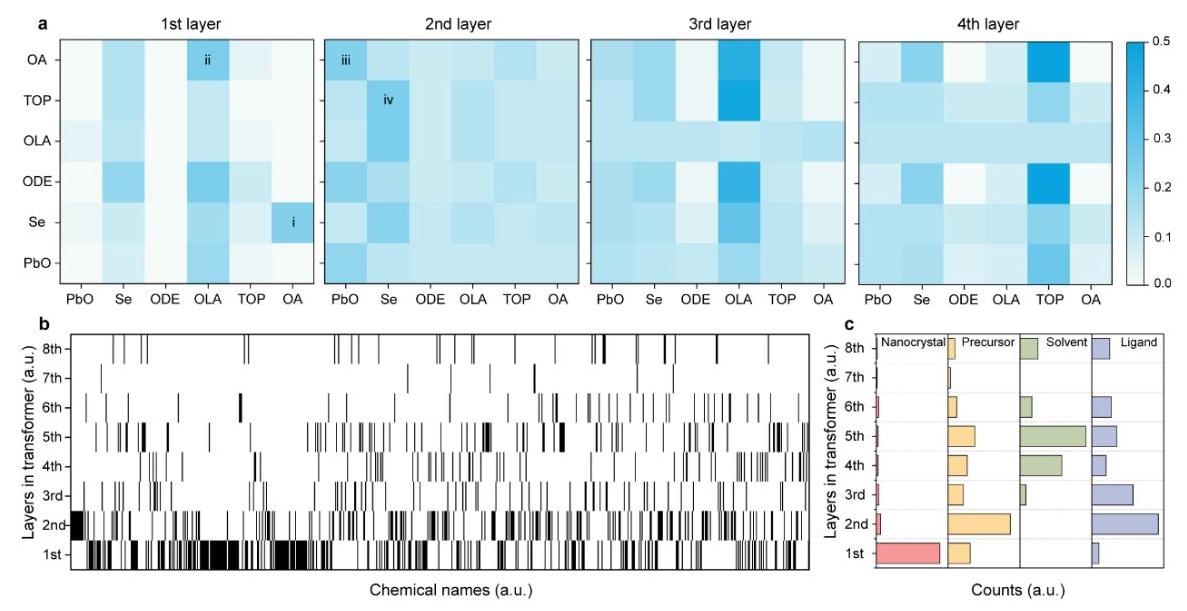}
    \caption*{Fig. 4 | Analysis of attention weights in the synthesis model. a, Heatmaps of the attention weights between chemicals in different layers. b, Maximum average attention weight maps of CLS for various chemicals across different transformer layers. The vertical lines indicate the maximum average attention weight of CLS for a given chemical, with the corresponding horizontal coordinate representing the chemical names and the vertical axis indicating the transformer layer. c, Categorization and distribution of highlighted chemicals across transformer layers.}
\end{figure}

In this work, we constructed a comprising 3500 synthesis recipes of 348 nanocrystal compositions and a dataset with 12000 TEM images. Based on the datasets, we developed a deep learning-based synthesis model by correlating the recipes with the sizes and shapes of nanocrystals. A semi-supervised learning algorithm was applied to derive the size and shapes of nanocrystals from TEM images, achieving an \({\mathrm{{AP}}}_{50}\) of \({82.5}\%\) . The synthesis model can achieve size prediction with an MAE of \({1.39}\mathrm{\;{nm}}\) and shape classification with average accuracy of \({89}\%\) . The generalizability tests show that the synthesis model can transfer knowledge across recipes of different nanocrystals, effectively predicting properties even when certain classes of nanocrystals are excluded from the dataset. Using the self-attention mechanism, the model can interpret the importance order of chemical reactions in size prediction. Through the analysis of the model, we gained insights into the importance of chemicals in determining nanocrystal synthesis. The nanocrystal synthesis model provides an excellent demonstration of data-driven approaches toward precise control of nanocrystals, which has potential to accelerate the developments of nanomaterials for industry applications.

\section*{Methods}

\subsection*{The generation of weak labels}

Nanocrystal segmentation models are typically trained on a limited set of expensive and manually labeled images without sufficiently utilizing large-scale unlabeled images. A semi-supervised learning method with a two-step weak label generation process was developed for training the segmentation model on unlabeled images, as shown in Extended Data Fig. 1.

\

\noindent\textit{Nanocrystal detection.} In the first step, a trained Cascade-RCNN \({}^{42}\) model is utilized to predict bounding boxes for nanocrystals in the unlabeled images. The primary challenge in nanocrystal detection lies in small and dense targets. Two strategies are adopted to improve detection accuracy: i) Using the Normalized Gaussian Wasserstein Distance (NWD) \({}^{43}\) instead of the Intersection over Union (IoU) metric. Specifically, the bounding boxes are modeled as 2D Gaussian distributions, with each bounding box determined by its center coordinates, size, and shape parameters. The IoU metric only considers the overlapping area of bounding boxes and is sensitive to minor positional deviations. The NWD considers the position, shape, and size of the bounding boxes to detect small and dense targets. ii) Using HRNet \({}^{44}\) as the backbone network. HRNet maintains high-resolution feature maps throughout the network and enhances the recognition ability of small targets through multi-scale fusion.

\

\noindent\textit{Uncertainty mask segmentation.} In the second step, each nanocrystal is subdivided into boundaries to generate the weak labels for training segmentation model on unlabeled images. Binary segmentation of each nanocrystal and the background within the predicted bounding box is performed using Otsu's Thresholding method to generate the mask image. Subsequently, erosion and dilation operations are implemented on these segmented mask images to delineate the boundaries of the segmentation masks. These boundaries are considered as regions of uncertainty in the segmentation. During the process of semi-supervised training, these uncertain regions are omitted from the loss calculation to avoid negatively impacting model performance. The regions outside these boundaries, which encompass relatively accurate foreground and background areas, are used for training the model on unlabeled images. Specifically, binary cross-entropy loss is applied to the foreground and background to improve the accuracy of segmentation model.

\subsection*{Nanocrystal segmentation model}

\noindent\textbf{TEM images dataset.} The segmentation model is first pretrained on a mixed labeled dataset and an unlabeled dataset, and then fine-tuned on our manually labeled dataset. This mixed dataset contains 4,880 images sourced from 12 different datasets, with totally 289,447 annotated instances (see Supplementary Table 1). The unlabeled dataset consists of 7,344 images with approximately 920,000 nanocrystals. After pretraining, the segmentation model undergoes further fine-tuning on our manually annotated dataset, consisting of 523 images and 49,976 nanocrystals. We chose another 80 images containing 6,827 annotated nanocrystals to serve as a validation set.

\

\noindent\textbf{Training segmentation model.} A neural network based on U-Net \({}^{45}\) is pretrained on the mixed labeled dataset and unlabeled dataset with generated weak labels using specially designed losses, as shown in Fig. 2. This neural network comprises an encoder and two decoders, producing a triplet of outputs: seed maps, horizontal gradients, and vertical gradients. These outputs together infer the final instance maps. For labeled images in the mixed dataset, ground truth labels for this triplet are derived from the full labels and employed for supervised training. Specifically, the seed map \({\mathbf{Y}}_{\mathbf{2}}\) indicates the pixel probabilities, supervised by the binary mask image \(\mathbf{P}\) indicating whether a given pixel is inside the nanocrystal target. This binary mask image \(\mathbf{P}\) can be generated from complete instance labels or weak labels. The use of seed maps is insufficient to differentiate between individual nanocrystals. An instance decoder is employed to predict and segment different nanocrystal boundaries. The instance decoder outputs horizontal gradients \({\mathbf{Y}}_{\mathbf{0}}\) and vertical gradients \({\mathbf{Y}}_{\mathbf{1}}\) , which are supervised by an Auxiliary vector flow representation computed from the complete instance labels. It transforms the full labels into true horizontal gradients \(\mathbf{H}\) and vertical gradients \(\mathbf{V}\) of the same size as the original image. These gradients are intended to locally transform pixels into other pixels within the cell and globally transform pixels into fixed points of the final gradient vector field over multiple iterations. These fixed points are selected as the centers of nanocrystals. Its horizontal and vertical gradients \(\mathbf{H}\) and \(\mathbf{V}\) represent the two vector fields of the Auxiliary vector flow representation. The loss function \(\mathrm{L}\) can be written as: \(\mathrm{L} = {\begin{Vmatrix}{\mathbf{Y}}_{\mathbf{0}} - \mathbf{H}\end{Vmatrix}}^{2} + {\begin{Vmatrix}{\mathbf{Y}}_{\mathbf{1}} - \mathbf{V}\end{Vmatrix}}^{2} + \mathrm{{BCE}}\left( {\sigma \left( {\mathbf{Y}}_{\mathbf{2}}\right) ,\mathrm{P}}\right)\) , where \(\sigma\) is the sigmoid function, and BCE is the binary cross-entropy. For unlabeled images, gradient labels cannot be calculated due to unclear boundaries of weak labels. Weak labels are solely used to supervise the seed maps \({\mathbf{Y}}^{\prime }{}_{2}\) predicted from unlabeled images, using binary masks \({\mathbf{P}}^{\prime }\) without uncertain regions that is not involved in the semi- supervised loss \({\mathrm{L}}_{\text{ semi }}\) calculation:

\[
{\mathrm{L}}_{\text{ sim }} = \mathrm{{BCE}}\left( {\sigma \left( {{\mathbf{Y}}^{\prime }2}\right) ,{\mathbf{P}}^{\prime }}\right) .
\]

\

\noindent\textbf{Implementation details.} The input images for the segmentation model are resized to a resolution of \({512} \times  {512}\) pixels. The pre-trained segmentation model is trained using stochastic gradient descent over 100 epochs, with a learning rate initially set at 0.2, momentum at 0.9, a batch size of 8, and a weight decay of 0.00001 . The learning rate begins at zero and linearly increases to 0.2 during the first 10 epochs to address initial training instability. Subsequent fine-tuning adopts the same learning rate strategy but extends over 200 epochs. To augment the training dataset, random spatial transformations are applied to both the training images and their corresponding labels, including random rotations, scaling, and translations, followed by the crop of a \({512} \times  {512}\) pixel segment from the center of the synthesized image. Additionally, ColorJitter is used to randomly adjust the brightness, contrast, saturation, and hue of the images, enhancing data robustness and model generalization. The model is trained on an Nvidia GeForce RTX 3090 GPU, achieving convergence in approximately six hours.

\

\noindent\textbf{Evaluation metrics.} In evaluating segmentation model, the mean Intersection over Union (mIOU) serves as a fundamental metric. This metric quantifies the overlap between the predicted segmentation and the ground truth by calculating the ratio of their intersection to their union. A predicted segmentation is considered accurate when its IOU with any ground truth instance surpasses a pre-defined threshold. Following the assessment of overlap accuracy, mean Average Precision (AP) at various thresholds is used to measure the precision of the model’s predictions. Specifically, \({\mathrm{{AP}}}_{50},{\mathrm{{AP}}}_{75}\) , and \({\mathrm{{AP}}}_{90}\) represent the average precision at IOU thresholds of0.50,0.75, and 0.90, respectively. The weights of the model that demonstrates the highest \({\mathrm{{AP}}}_{50}\) on the validation set during training are preserved.

\

\noindent\textbf{Unsupervised shape clustering.} Unsupervised shape clustering is employed as a robust method to examine the distribution and variance in physical geometry under diverse synthesis conditions. Several shape descriptors are extracted from the instance masks generated by our segmentation model, including solidity, convexity, eccentricity, aspect ratio, and circularity. The distributions of these descriptors are illustrated in Supplementary Fig. 2. K-means clustering and t-distributed stochastic neighbor embedding are applied to shape descriptors to categorize and visualize the impact of synthesis parameters on shape evolution.

\

\noindent\textbf{Calculation of equivalent circle diameter.} Considering the differences in the shapes of nanocrystals, we defined the equivalent circle diameter to compare their sizes. Masks for each nanocrystal in the image are obtained by the segmentation model, enabling to calculate the pixel area \(S\) . Then, EasyOCR is employed to identify the numerical value and units of the scale bar in TEM images, providing the true length of the scale bar \({l}_{\text{ true }}\) . The pixel length \({l}_{\text{ pixel }}\) of the scale bar is then obtained using a simple global thresholding strategy. Therefore, the equivalent circle diameter \(d\) of the nanocrystal can be calculated using the formula: \(d = \frac{{l}_{\text{ ture }}}{{l}_{\text{ pixel }}}\sqrt{\frac{4S}{\pi }}\) . Therefore, the size labels \(D\) for each recipe can be calculated as: \(D = \left( {{d}_{1} + {d}_{2} + \ldots  + {d}_{N}}\right) /N\) , where \(N\) is the statistical number of nanocrystals.

\subsection*{Nanocrystal synthesis model}

\noindent\textbf{Construction of chemical descriptors.} The construction process of chemical descriptors is shown in Extended Data Fig. 5a. Firstly, the names or formulas of chemicals are extracted from the recipe dataset. The 3D structures of all crystalline chemicals are acquired from the open-access Materials Project database \({}^{46}\) . While, the 3D structures of organic molecules are obtained through DFT calculations performed in Materials Studio (DMol \({}^{3}\) module) \({}^{47}\) . The generalized gradient approximation in the Perdew-Burke-Ernzerhof form with the double numerical basis sets plus the polarization functional are adopted \({}^{{48},{49}}\) . The density mixing fraction of 0.2 with direct inversion in the iterative subspace and orbital occupancy with smearing of \({0.005}\mathrm{\;{Ha}}\) are employed. The self-consistent field tolerance is set to \({10}^{-6}\mathrm{\;{Ha}}/\) atom. The total energy, maximum force and maximum displacement are set as \({10}^{-6}\mathrm{\;{Ha}}/\) atom, \({0.002}\mathrm{\;{Ha}}/Å\) and \({0.005Å}\) , respectively for geometry optimization. Subsequently, a pre-trained GNN model \({}^{50}\) is fine-tuned using 1,797 chemical structures in our recipes to extract the chemical descriptors. These descriptors are represented as 512-dimensional features, providing a comprehensive numerical representation of the structural characteristics.

\

\noindent\textbf{Pipeline.} Given the variable number of chemicals in different recipes, the transformer algorithm \({}^{51}\) (refer to Supplementary Fig. 5) is utilized to train the synthesis model for the size and shape prediction. The transformer is particularly adept at handling unordered, variable-length input data and employ self-attention mechanisms to discern relationships among chemicals in sequences of any length. This capability is vital for understanding and predicting the interactions among chemicals and their influence on the final chemical reaction outcomes.

\

\noindent\textit{Input preprocessing.} For the transformer's input, chemical descriptors are amalgamated with reaction descriptors to create a feature representation sequence for the various materials in a given recipe. Reaction descriptors include injection temperature \({T}_{\mathrm{j}}\) , reaction temperature \({T}_{\mathrm{r}}\) , reaction time \(t\) , heating rate \({Sp}\) , and the molar quantities \({Mol}\) corresponding to the chemicals. Considering the uneven distribution of reaction conditions and molar quantities (Supplementary Fig. 8), a robust scaler is employed for standardization, scaling features using statistics that are resilient to outliers. Subsequently, a learnable CLS token is incorporated into the input sequence as a descriptor for chemical reactions. This token captures global contextual information of the entire sequence and forms the foundation for the final output predictions. The sequence, including the CLS token, undergoes a linear projection before entering the multi-layer transformer architecture.

\

\noindent\textit{Transformer layers pass.} For each layer, layer normalization is performed at the beginning, stabilizing the training of deep neural networks. It normalizes the inputs across the features instead of across the batch, and this is done for each individual sample. It accelerates the training and converging quicker by reducing the internal covariate shift. After normalization, the transformer processes the data through a multihead self-attention mechanism. Each input element is transformed into three vectors: queries (Q), keys (K), and values (V). Self-Attention mechanism allows a model to weigh the relevance of different parts of an input sequence independently of their position in the sequence. These vectors are computed by multiplying the input embeddings by three learned weight matrices specific to queries, keys, and values, respectively. In multi-head self-attention, the model applies multiple sets of \(\mathbf{Q},\mathbf{K}\) , and \(\mathbf{V}\) matrices, each representing a different ``head". Each head can potentially learn to attend to different parts of the input sequence, capturing various aspects of the data. After self-attention, position-wise feed-forward network applies a set of linear transformations to each element separately and identically. The feed-forward network consists of two linear transformations with a ReLU fiction in between. This module can enhance the representation capacity of the transformer without considering the sequence's positional order, focusing instead on transforming the features. Finally, similar to the post self-attention phase, the output from the feed-forward network is added back to the input of the feed-forward network itself, i.e., ``Add \& Norm" operator. This addition is again followed by layer normalization. Finally, the features of the CLS token after the transformer are concatenated with the reaction descriptors containing \({T}_{\mathrm{j}}\) , \({T}_{\mathrm{r}},t\) , and \({Sp}\) , and fed into linear layers with ReLU activation functions to make the final predictions regarding the size and shape of the nanocrystals.

\

\noindent\textbf{Data augmentation.} To overcome the challenge of limited training data, a novel data augmentation method is introduced based on reaction intermediates (Supplementary Fig. 5b). This method significantly enhances both the quantity and diversity of data available for model training. The foundational principle of this approach allows for any two chemicals within a recipe to react. For example, when considering PbO and OA, they react (assuming a molar ratio of 1:1) to generate an intermediate chemical, PbO-OA. The 3D structure of this intermediate is derived through DFT calculations. This structure is then input into our fine-tuned GNN to obtain its descriptor. Subsequently, the descriptor of PbO-OA is incorporated into the sequence, and the molar quantities of \(\mathrm{{PbO}},\mathrm{{OA}}\) , and \(\mathrm{{PbO}} - \mathrm{{OA}}\) are updated, resulting in the generation of a new recipe. More specifically, probabilities are set such that there is a 50\% chance to simulate a complete reaction of the materials and another 50\% chance to simulate a partial reaction. The occurrence of one or multiple intermediates simultaneously is managed by manipulating random numbers. Through this approach, each recipe can theoretically be augmented into several dozen new recipes featuring different intermediates and molar quantities. With generated augmentation recipes, the model is pretrained on both the original and augmented data to enable the model to learn the entire process from chemicals to intermediates to final products. Then, the model is fine-tuned using only the original data to ensure its accuracy and generalization capabilities when predicting actual recipes. To better adapt the model to the complexity of real reactions with more diverse data, negative recipes (unsuccessful preparation of nanocrystals) are collected from experimental records and literature. These recipes do not have size and shape labels. This problem is addressed by adding an extra output to the model by setting a binary classification output to distinguish whether the reaction success with formed nanocrystals and use binary cross-entropy loss for training.

\

\noindent\textbf{Training synthesis model.} For shape classification, the cross-entropy loss function is used for training. For size prediction, the Huber loss function is used for training due to the broad range and uneven distribution of size labels (Supplementary Fig. 7). The Huber loss function is a robust alternative to other loss functions, such as the mean squared error and absolute error functions. It is designed to provide a balance between accuracy and sensitivity, particularly in situations where there are large errors in size prediction. Considering the statistical error in size, a tolerance threshold of \(\pm  {15}\%\) is also set for datasets containing fewer than 300 nanocrystals. If the relative error between the model's predicted size and the actual size label falls within this range, their losses are set to zero. This approach is designed to mitigate prediction errors potentially arising from the model overly fitting to a small amount of inaccurate data.

\

\noindent\textbf{Performance evaluation.} Inaccuracy of size labels extracted from literature due to small number of manual measurements of nanocrystals. Therefore, a five-fold cross-validation method is employed for data with nanocrystal counts greater than \({300}(N >\) 300). The model's performance is then evaluated by training and validating across all five folds, and reporting the average performance. In each round of validation, four subsets, along with additional data containing fewer than 300 nanocrystals for statistics, are used for training, while the remaining subset serves as the validation set. For shape classification, accuracy metrics are reported. For size regression, various statistical measures are assessed, including mean absolute error (MAE), mean squared error (MSE), root mean squared error (RMSE), mean absolute percentage error (MAPE), and the coefficient of determination \(\left( {\mathrm{R}}^{2}\right)\) . The results are as shown in Extended Data Table

\

\noindent\textbf{Generalizability evaluation.} Considering that the diversity of synthesis recipes that contains a huge potential space, we designed several generalizability tests to evaluate the performance of the model for given new chemicals or chemical combinations. These tests involve selecting the whole recipes for one type of nanocrystal to serve as the validation set, while using the remaining data for training. This test is actually quite rigorous because the chemicals and combinations of chemicals in the validation set may be completely new. This approach allows us to test whether our model truly learns to generalize and make inferences from the data. Four training and validation sessions are conducted, including four representative types of nanocrystals: PbSe, \({\mathrm{{Ag}}}_{2}\mathrm{\;S},\mathrm{{Ni}}\) , and \({\mathrm{{Cs}}}_{2}{\mathrm{{NaBiCl}}}_{6}\) . By employing this dataset division and validation set selection strategy, a more comprehensively and accurately evaluation can be done to show the potential of our model's generalizability in predicting the size of new nanocrystals, as shown in Extended Data Table 2.

\

\noindent\textbf{Attention weight visualization.} In the transformer model, the attention weights of each chemical in the input sequence are computed with respect to every other chemical in each self-attention layer. These weights represent the importance of each chemical to the model when processing the input information. For each chemical feature in the sequence, the model generates three types of vectors through different linear transformations: \(\mathbf{Q},\mathbf{K}\) , and \(\mathbf{V}\) . The attention weights are calculated between the \(\mathbf{Q}\) and all \(\mathbf{K}\) , followed by softmax normalization to scale the weights appropriately:

\[
\operatorname{Attention}\left( {\mathbf{Q},\mathbf{K},\mathbf{V}}\right)  = \operatorname{softmax}\left( \frac{{\mathbf{{QK}}}^{\mathrm{T}}}{\sqrt{{d}_{\mathrm{k}}}}\right) \mathbf{V}
\]

where \({d}_{\mathrm{k}}\) is the dimension of vectors. Through this method, the self-attention mechanism dynamically focuses on the importance of different parts of the input sequence. Then, these weights are extracted from the trained model. The attention degree of each chemical in different transformer layers is defined as the average attention weight between the CLS token and the chemical:

\[
{\mathbf{A}}_{o,l} = \mathop{\sum }\limits_{{m = 1}}^{M}\operatorname{softmax}\left( \frac{{\mathbf{Q}}_{\mathrm{{CLS}},l}{\mathbf{K}}_{o,l}^{\mathrm{T}}}{\sqrt{{d}_{\mathrm{k}}}}\right) /M
\]

where \({\mathrm{A}}_{\mathrm{o},1}\) denotes the attention degree of material \(o\) at layer \(l\) , and \(M\) represents the number of reactions involving material \(o.{\mathbf{Q}}_{\mathrm{{CLS}},l}\) is the query vector of the CLS token at layer \(l\) for a given recipe, while \({\mathbf{K}}^{\mathrm{T}}{}_{o,l}\) is the transpose of the key vector for material \(o\) at the same layer. Supplementary Fig. 6 visualizes the attention each material receives across different transformer layers. Fig. 4b highlights the layers where each material receives the maximum attention, pinpointing where significant interactions occur in the network. This visualization technique shows the model's focus across different stages of processing in the transformer model, helping to understand how it interprets and prioritizes information in chemical reaction prediction.

\section*{Code availability}

TEM image dataset and algorithm for nanocrystal segmentation is available at \url{https://github.com/Sharpiless/Nanocrystals-TEM-segmentation}. Recipe dataset and algorithm for nanocrystal synthesis is available at \url{https://github.com/Sharpiless/Nanocrystals-Deep-Learning}.

\section*{References}

\begin{enumerate}
    \item García de Arquer, F. P. et al. Semiconductor quantum dots: Technological progress and future challenges. \textit{Science} \textbf{373}, eaaz8541 (2021).
    \item Kagan, C. R., Lifshitz, E., Sargent, E. H. \& Talapin, D. V. Building devices from colloidal quantum dots. \textit{Science} \textbf{353}, aac5523 (2016).
    \item Niu, Z. et al. Anisotropic phase segregation and migration of Pt in nanocrystals en route to nanoframe catalysts. \textit{Nat. Mater.} \textbf{15}, 1188–1194 (2016).
    \item Li, X.-B., Tung, C.-H. \& Wu, L.-Z. Semiconducting quantum dots for artificial photosynthesis. \textit{Nat. Rev. Chem.} \textbf{2}, 160–173 (2018).
    \item Jin, D. et al. Nanoparticles for super-resolution microscopy and single-molecule tracking. \textit{Nat. Methods} \textbf{15}, 415–423 (2018).
    \item Alivisatos, A. P. Semiconductor clusters, nanocrystals, and quantum dots. \textit{Science} \textbf{271}, 933–937 (1996).
    \item Norris, D. J. \& Bawendi, M. Measurement and assignment of the size-dependent optical spectrum in CdSe quantum dots. \textit{Phys. Rev. B} \textbf{53}, 16338 (1996).
    \item Scholes, G. D. \& Rumbles, G. Excitons in nanoscale systems. \textit{Nat. Mater.} \textbf{5}, 683–696 (2006).
    \item Jun, Y.-w., Choi, J.-s. \& Cheon, J. Shape control of semiconductor and metal oxide nanocrystals through nonhydrolytic colloidal routes. \textit{Angew. Chem. Int. Ed.} \textbf{45}, 3414–3439 (2006).
    \item Kovalenko, M. V. et al. Prospects of nanoscience with nanocrystals. \textit{ACS Nano} \textbf{9}, 1012–1057 (2015).
    \item Yarema, O., Yarema, M. \& Wood, V. Tuning the composition of multicomponent semiconductor nanocrystals: The case of I–III–VI materials. \textit{Chem. Mater.} \textbf{30}, 1446–1461 (2018).
    \item Efros, A. L. \& Brus, L. E. Nanocrystal quantum dots: From discovery to modern development. \textit{ACS Nano} \textbf{15}, 6192–6210 (2021).
    \item Long, Z. et al. A reactivity-controlled epitaxial growth strategy for synthesizing large nanocrystals. \textit{Nat. Synth.} \textbf{2}, 296–304 (2023).
    \item Li, J. J. et al. Large-scale synthesis of nearly monodisperse CdSe/CdS core/shell nanocrystals using air-stable reagents via successive ion layer adsorption and reaction. \textit{J. Am. Chem. Soc.} \textbf{125}, 12567–12575 (2003).
    \item Tessier, M. D., Dupont, D., De Nolf, K., De Roo, J. \& Hens, Z. Economic and size-tunable synthesis of InP/ZnE (E = S, Se) colloidal quantum dots. \textit{Chem. Mater.} \textbf{27}, 4893–4898 (2015).
    \item Pu, Y., Cai, F., Wang, D., Wang, J.-X. \& Chen, J.-F. Colloidal synthesis of semiconductor quantum dots toward large-scale production: A review. \textit{Ind. Eng. Chem. Res.} \textbf{57}, 1790–1802 (2018).
    \item Lee, E., Wang, C., Yurek, J. \& Ma, R. A new frontier for quantum dots in displays. \textit{Inf. Disp.} \textbf{34}, 10–31 (2018).
    \item Liu, M. et al. Probing intermediates of the induction period prior to nucleation and growth of semiconductor quantum dots. \textit{Nat. Commun.} \textbf{8}, 15467 (2017).
    \item van Embden, J., Gross, S., Kittilstved, K. R. \& Della Gaspera, E. Colloidal approaches to zinc oxide nanocrystals. \textit{Chem. Rev.} \textbf{123}, 271–326 (2023).
    \item Thanh, N. T. K., Maclean, N. \& Mahiddine, S. Mechanisms of nucleation and growth of nanoparticles in solution. \textit{Chem. Rev.} \textbf{114}, 7610–7630 (2014).
    \item Akkerman, Q. A. et al. Controlling the nucleation and growth kinetics of lead halide perovskite quantum dots. \textit{Science} \textbf{377}, 1406–1412 (2022).
    \item Sanchez-Lengeling, B. \& Aspuru-Guzik, A. Inverse molecular design using machine learning: Generative models for matter engineering. \textit{Science} \textbf{361}, 360–365 (2018).
    \item Xie, T. \& Grossman, J. C. Crystal graph convolutional neural networks for an accurate and interpretable prediction of material properties. \textit{Phys. Rev. Lett.} \textbf{120}, 145301 (2018).
    \item Peng, J., Muhammad, R., Wang, S.-L. \& Zhong, H.-Z. How machine learning accelerates the development of quantum dots? \textit{Chin. J. Chem.} \textbf{39}, 181–188 (2021).
    \item Deng, B. et al. CHGNet as a pretrained universal neural network potential for charge-informed atomistic modelling. \textit{Nat. Mach. Intell.} \textbf{5}, 1031–1041 (2023).
    \item Szymanski, N. J. et al. An autonomous laboratory for the accelerated synthesis of novel materials. \textit{Nature} \textbf{624}, 86–91 (2023).
    \item Slattery, A. et al. Automated self-optimization, intensification, and scale-up of photocatalysis in flow. \textit{Science} \textbf{383}, eadj1817 (2024).
    \item Abolhasani, M. \& Kumacheva, E. The rise of self-driving labs in chemical and materials sciences. \textit{Nat. Synth.} \textbf{2}, 483–492 (2023).
    \item Tao, H. et al. Nanoparticle synthesis assisted by machine learning. \textit{Nat. Rev. Mater.} \textbf{6}, 701–716 (2021).
    \item Masson, J.-F., Biggins, J. S. \& Ringe, E. Machine learning for nanoplasmonics. \textit{Nat. Nanotechnol.} \textbf{18}, 111–123 (2023).
    \item Voznyy, O. et al. Machine learning accelerates discovery of optimal colloidal quantum dot synthesis. \textit{ACS Nano} \textbf{13}, 11122–11128 (2019).
    \item Baum, F., Pretto, T., Köche, A. \& Santos, M. J. L. Machine learning tools to predict hot injection syntheses outcomes for II–VI and IV–VI quantum dots. \textit{J. Phys. Chem. C} \textbf{124}, 24298–24305 (2020).
    \item Nguyen, H. A. et al. Predicting indium phosphide quantum dot properties from synthetic procedures using machine learning. \textit{Chem. Mater.} \textbf{34}, 6296–6311 (2022).
    \item Pyrz, W. D. \& Buttrey, D. J. Particle size determination using TEM: A discussion of image acquisition and analysis for the novice microscopist. \textit{Langmuir} \textbf{24}, 11350–11360 (2008).
    \item Lee, B. et al. Statistical characterization of the morphologies of nanoparticles through machine learning based electron microscopy image analysis. \textit{ACS Nano} \textbf{14}, 17125–17133 (2020).
    \item Yildirim, B. \& Cole, J. M. Bayesian particle instance segmentation for electron microscopy image quantification. \textit{J. Chem. Inf. Model.} \textbf{61}, 1136–1149 (2021).
    \item Lin, B. et al. A deep learned nanowire segmentation model using synthetic data augmentation. \textit{npj Comput. Mater.} \textbf{8}, 88 (2022).
    \item Paik, T. et al. Shape-controlled synthesis and self-assembly of highly uniform upconverting calcium fluoride nanocrystals. \textit{Inorg. Chem. Front.} \textbf{11}, 278–285 (2024).
    \item Lundberg, S. \& Lee, S.-I. A unified approach to interpreting model predictions. Preprint at \textit{arXiv:1705.07874} (2017).
    \item Mourdikoudis, S. et al. Oleic acid/oleylamine ligand pair: a versatile combination in the synthesis of colloidal nanoparticles. \textit{Nanoscale Horiz.} \textbf{7}, 941–1015 (2022).
    \item Preske, A. et al. Size-programmed synthesis of PbSe quantum dots via secondary phosphine chalcogenides. \textit{Chem. Mater.} \textbf{31}, 8301–8307 (2019).
    \item Cai, Z. \& Vasconcelos, N. Cascade R-CNN: Delving into high quality object detection. In \textit{2018 IEEE/CVF Conference on Computer Vision and Pattern Recognition} 6154–6162 (IEEE, 2018).
    \item Wang, J., Xu, C., Yang, W. \& Yu, L. A normalized gaussian wasserstein distance for tiny object detection. Preprint at \textit{arXiv:2110.13389} (2021).
    \item Wang, J. et al. Deep high-resolution representation learning for visual recognition. \textit{IEEE Trans. Pattern Anal. Mach. Intell.} \textbf{43}, 3349–3364 (2021).
    \item Stringer, C., Wang, T., Michaelos, M. \& Pachitariu, M. Cellpose: a generalist algorithm for cellular segmentation. \textit{Nat. Methods} \textbf{18}, 100–106 (2021).
    \item Jain, A. et al. Commentary: The Materials Project: A materials genome approach to accelerating materials innovation. \textit{APL Mater.} \textbf{1}, (2013).
    \item Delley, B. An all-electron numerical method for solving the local density functional for polyatomic molecules. \textit{J. Chem. Phys.} \textbf{92}, 508–517 (1990).
    \item Perdew, J. P. et al. Atoms, molecules, solids, and surfaces: Applications of the generalized gradient approximation for exchange and correlation. \textit{Phys. Rev. B} \textbf{46}, 6671–6687 (1992).
    \item Perdew, J. P., Burke, K. \& Ernzerhof, M. Generalized gradient approximation made simple. \textit{Phys. Rev. Lett.} \textbf{77}, 3865–3868 (1996).
    \item Lu, S., Gao, Z., He, D., Zhang, L. \& Ke, G. Data-driven quantum chemical property prediction leveraging 3D conformations with Uni-Mol+. \textit{Nat. Commun.} \textbf{15}, 7104 (2024).
    \item Vaswani, A. et al. Attention is all you need. In \textit{Proceedings of the 31st International Conference on Neural Information Processing Systems} 6000–6010 (NIPS, 2017).
\end{enumerate}

\newpage

\section*{Acknowledgements}

This work was supported by National Natural Science Foundation of China (U23A20683, H.Z.) and Beijing Natural Science Foundation (Z210018, H.Z.). We thank Mingrui Liu, Zhiwei Long, and Shipei Sun for their contributions to the recipe dataset.

\section*{Supplementary Information}

\begin{figure}[htbp]
    \centering
    \includegraphics[max width=1.0\textwidth]{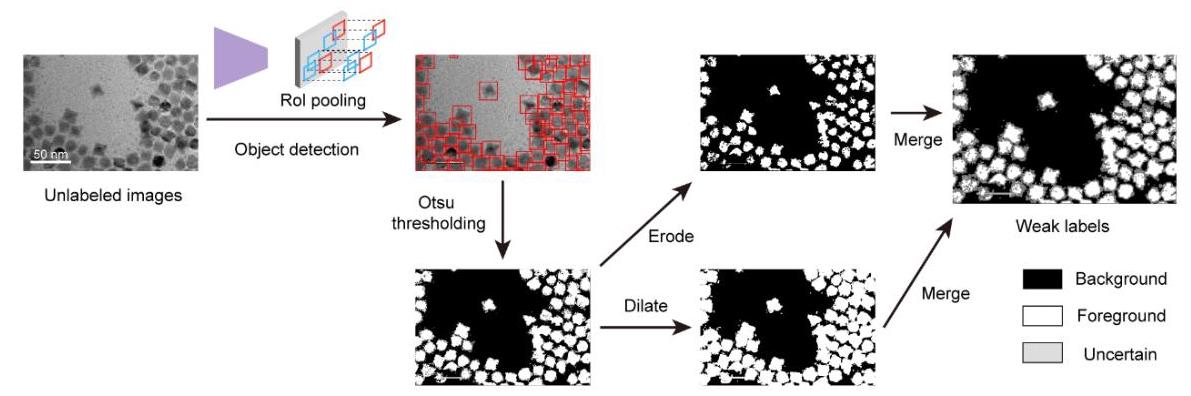}
    \caption*{Extended Data Fig. 1 | Workflow for generating weak labels from unlabeled images by a pre-trained object detection network and morphological operations.}
\end{figure}

\begin{figure}[htbp]
    \centering
    \includegraphics[max width=0.8\textwidth]{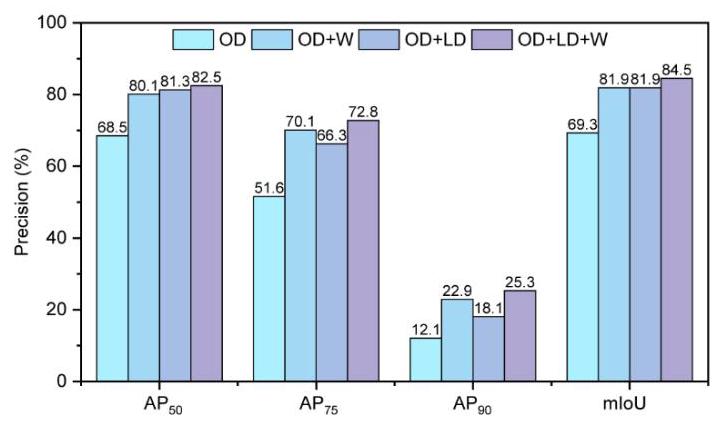}
    \caption*{Extended Data Fig. 2 | Performance comparison of segmentation models trained on different combinations of full labels, full labels from literature and weak labels. OD represents the use of full labels from our dataset; OD+W represents the use of both full and weak labels; OD+LD represents the use of full labels from our dataset and full labels from literature; OD+LD+W represents the use of the three types of labels. Evaluation metrics \(\left( {{\mathrm{{AP}}}_{50},{\mathrm{{AP}}}_{75},{\mathrm{{AP}}}_{90}}\right.\) , and \(\left. \mathrm{{mIoU}}\right)\) are defined in the Methods section. 14 The dataset sizes are provided in Supplementary Table 1.}
\end{figure}

\begin{figure}[htbp]
    \centering
    \includegraphics[max width=1.0\textwidth]{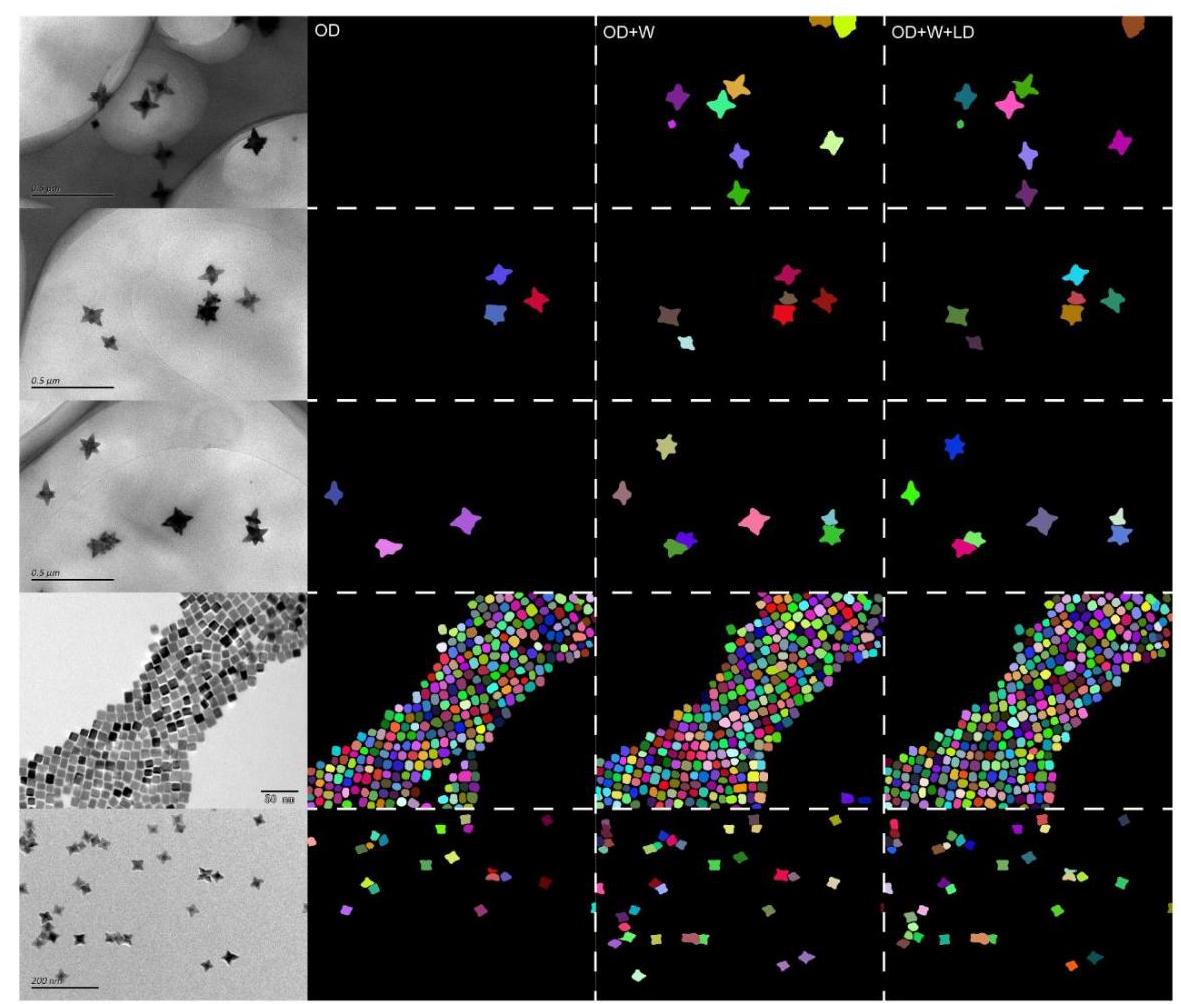}
    \caption*{Extended Data Fig. 3 | Segmentation maps of nanocrystals generated by models trained with various combinations of full labels, full labels from literature and weak labels. Segmentation results of nanocrystals using models trained with different datasets: OD (full labels), OD+W (full and weak labels), and OD+W+LD (full labels, weak labels, and full labels from literature). The incorporation of weak labels (third column) improves the model's recall and localization accuracy, while adding full labels from literature (fourth column) further refines the segmentation performance.}
\end{figure}

\begin{figure}[htbp]
    \centering
    \includegraphics[max width=1.0\textwidth]{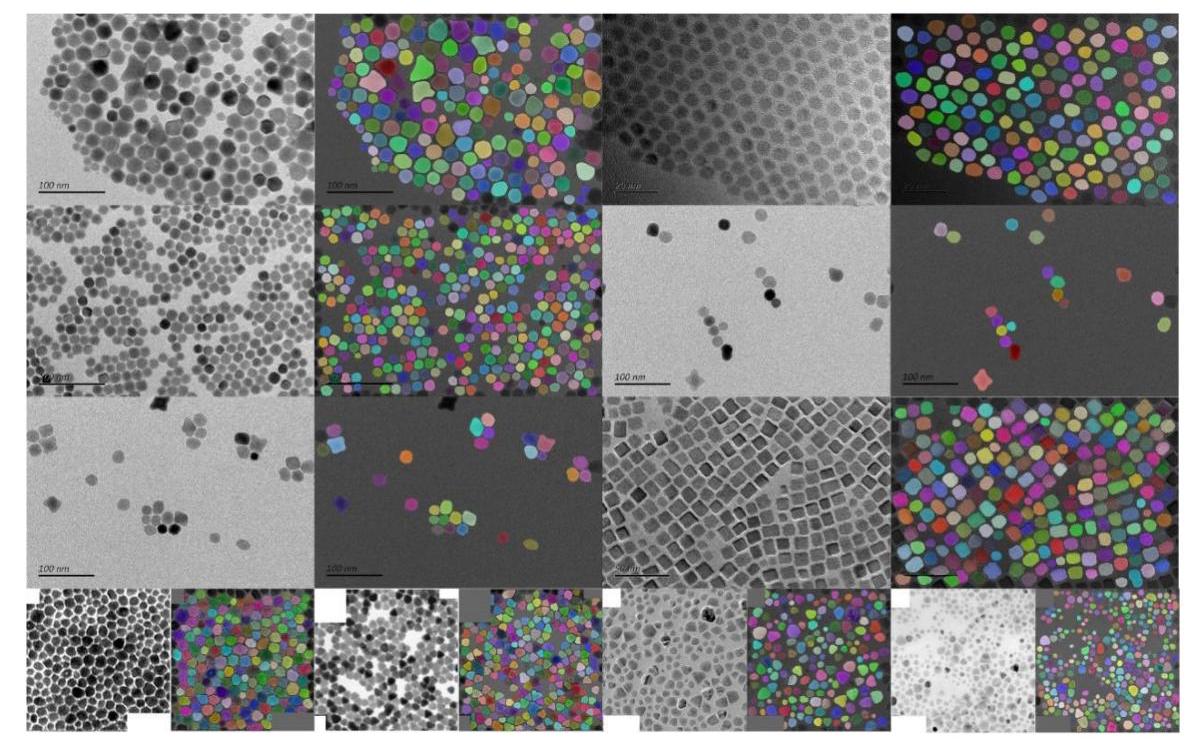}
    \caption*{Extended Data Fig. 4 | Comparison of TEM images of nanocrystals before and after segmentation. Nanocrystals of varying densities and shapes are effectively segmented. The final row of images consists of low-resolution images.}
\end{figure}

\begin{figure}[htbp]
    \centering
    \includegraphics[max width=1.0\textwidth]{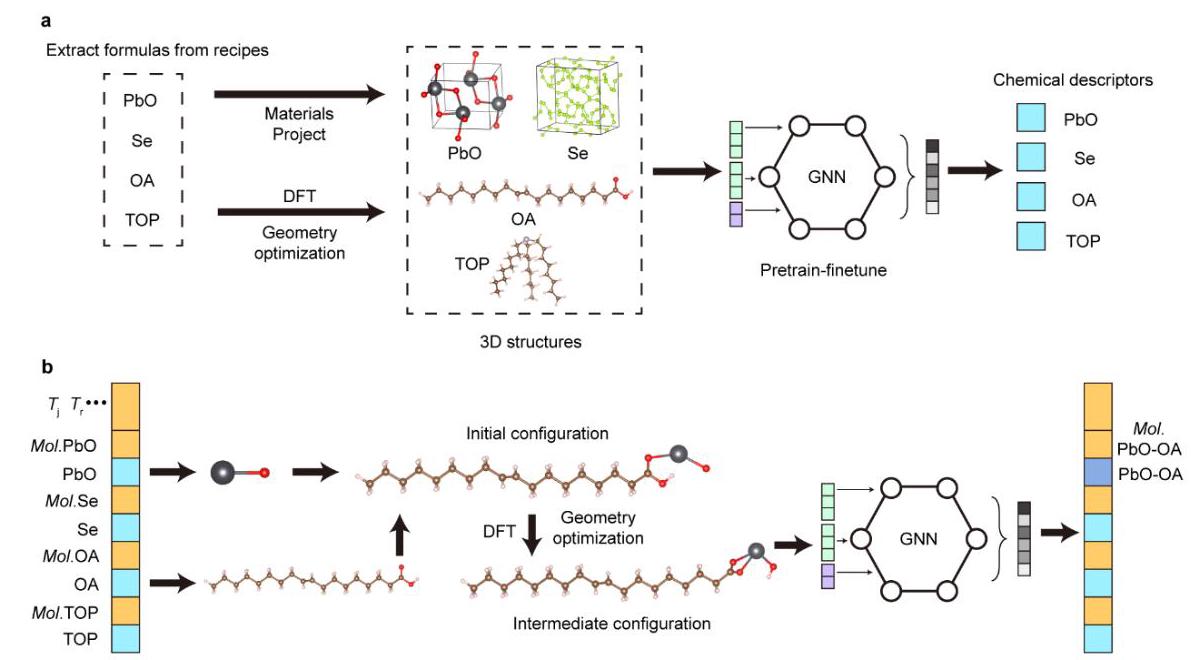}
    \caption*{Extended Data Fig. 5 | Workflow for generating chemical descriptors and implementing reaction intermediate-based data augmentation. a, Process for generating chemical descriptors: 3D structures of reactants are obtained from the Materials Project database or via DFT calculations. Chemical descriptors are obtained by using these 3D structures as input to the GNN. b, Reaction intermediate-based data augmentation method: Intermediates of reactants are calculated using DFT, followed by geometry optimization. Chemical descriptors for the intermediates are obtained via GNN, which are then used to update the original deep learning model inputs.}
\end{figure}

\begin{figure}[htbp]
    \centering
    \includegraphics[max width=0.9\textwidth]{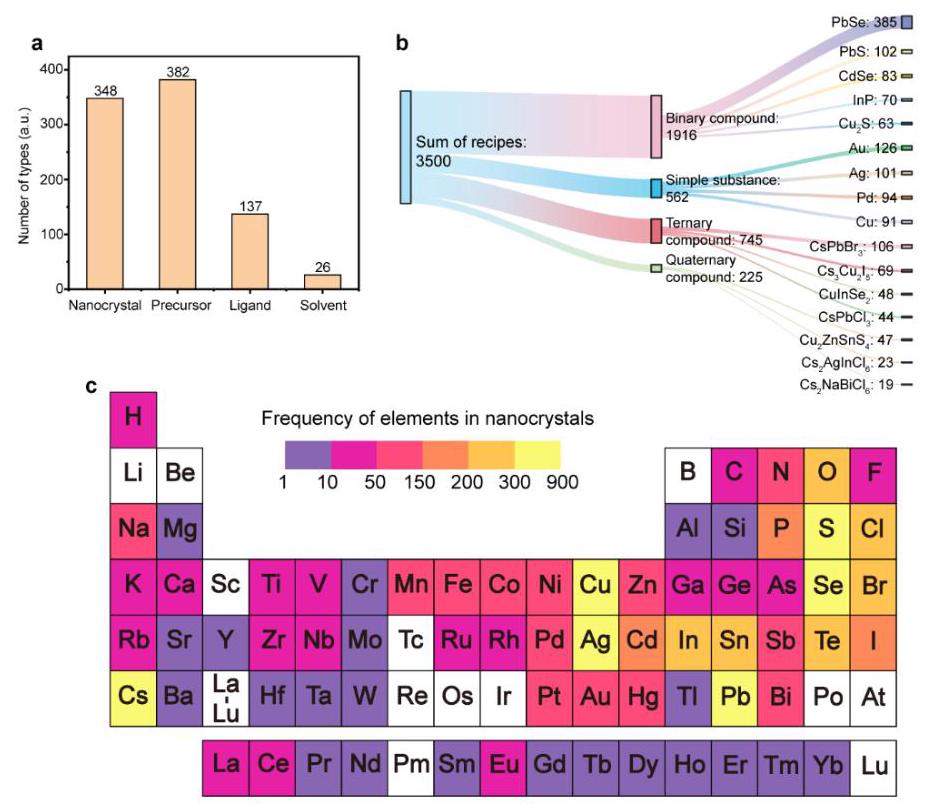}
    \caption*{Extended Data Fig. 6 | Distribution of the recipe dataset. a, Number of different types of chemicals included across all recipes, categorized as nanocrystals, precursors, ligands, and solvents. b, Distribution of recipes for the synthesis of various types of nanocrystals, categorized into simple substances, binary compounds, ternary compounds, and quaternary compounds. Specific examples and the number of recipes for each type are indicated. c, Elemental frequency distribution of different nanocrystals in the recipe dataset.}
\end{figure}

\begin{figure}[htbp]
    \centering
    \includegraphics[max width=0.8\textwidth]{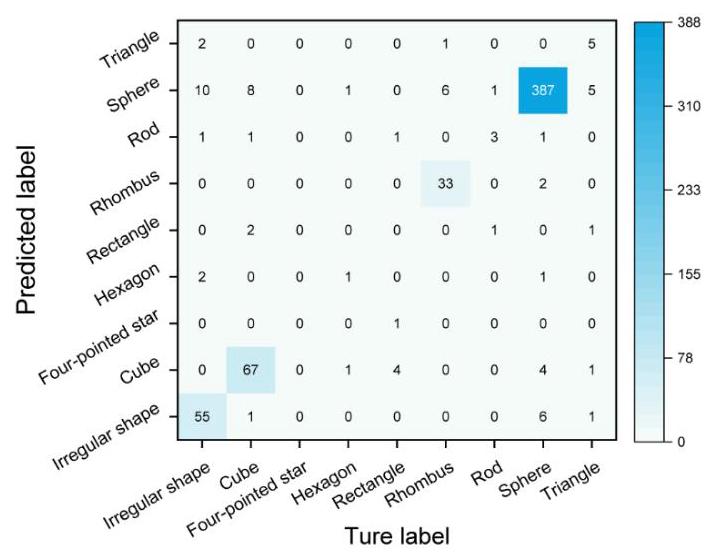}
    \caption*{Extended Data Fig. 7 | Confusion matrix for evaluating the transformer-trained model's performance in shape classification. The color intensity reflects the number 2 of instances for each prediction-outcome pair.}
\end{figure}

\begin{figure}[htbp]
    \centering
    \includegraphics[max width=0.8\textwidth]{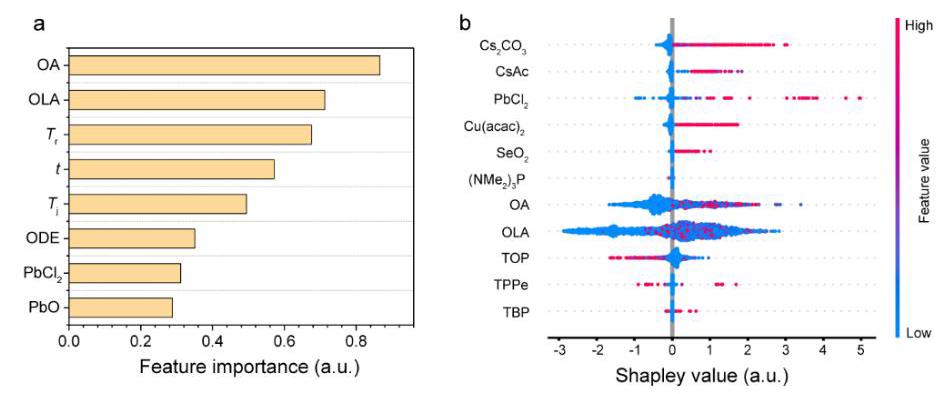}
    \caption*{Extended Data Fig. 8 | Shapley value analysis. a, Feature importance rankings calculated from all PbSe synthesis recipes, where OA, OLA, ODE, PbO and PbCl2 represent the molar amount of oleic acid, oleylamine, octadecene, lead chloride and lead oxide, respectively. \({T}_{\mathrm{r}},t\) and \({T}_{\mathrm{j}}\) represent reaction temperature, reaction time and injection temperature, respectively. b, Beeswarm plot of the chemical molar amounts in the recipe dataset for their contribution to nanocrystal size. \({\mathrm{{Cs}}}_{2}{\mathrm{{CO}}}_{3},\mathrm{{CsAc}},\mathrm{{Cu}}{\left( \mathrm{{acac}}\right) }_{2}\) , \({\mathrm{{SeO}}}_{2},{\left( {\mathrm{{NMe}}}_{2}\right) }_{3}\mathrm{P}\) , TOP, TPPe and TBP represent molar amounts of cesium carbonate, cesium acetate, copper acetylacetonate, selenium oxide, tris(dimethylamino)phosphine, tri-n-octylphosphine, triphenylphosphine and tributylphosphine, respectively. Each point represents a recipe containing the chemical, with the horizontal coordinate showing the Shapley value. A larger Shapley value indicates a greater contribution to the nanocrystal size. The color of each point reflects the molar amount of the chemical.}
\end{figure}

\begin{figure}[htbp]
    \centering
    \caption*{Extended Data Table 1 | Performance comparison of synthesis models trained with different algorithms for size prediction.}
    \includegraphics[max width=0.8\textwidth]{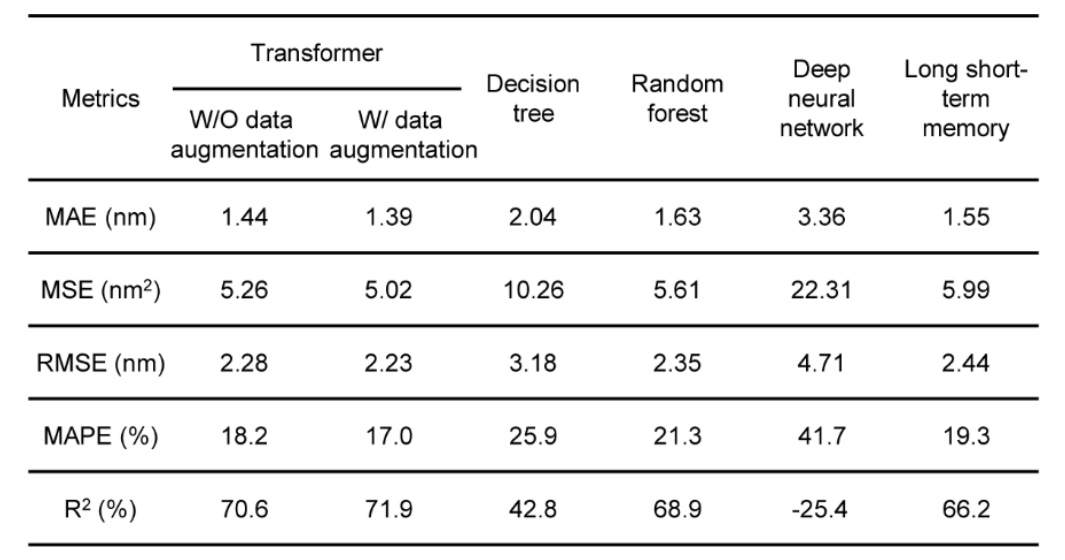}
\end{figure}

\begin{figure}[htbp]
    \centering
    \caption*{Extended Data Table 2 | Generalizability tests for size prediction.}
    \includegraphics[max width=1.0\textwidth]{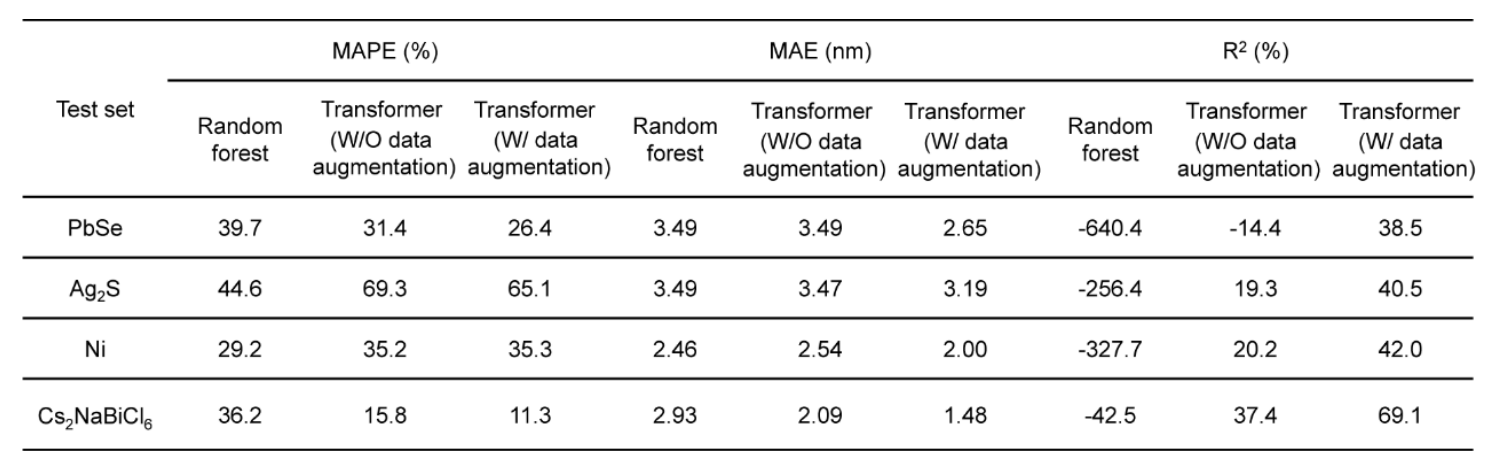}
\end{figure}

\begin{figure}[htbp]
    \centering
    \caption*{Supplementary Table 1 | Scale of the image datasets.}
    \includegraphics[max width=1.0\textwidth]{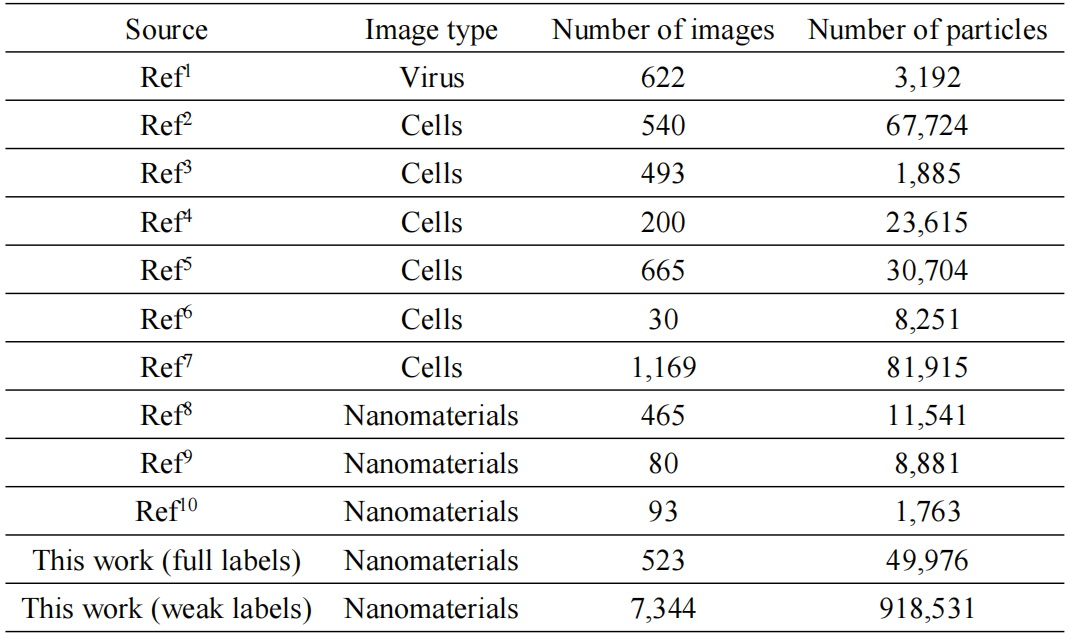}
\end{figure}

\begin{figure}[htbp]
    \centering
    \caption*{Supplementary Table 2 | Performance comparison of synthesis models trained with different algorithms for shape classification.}
    \includegraphics[max width=1.0\textwidth]{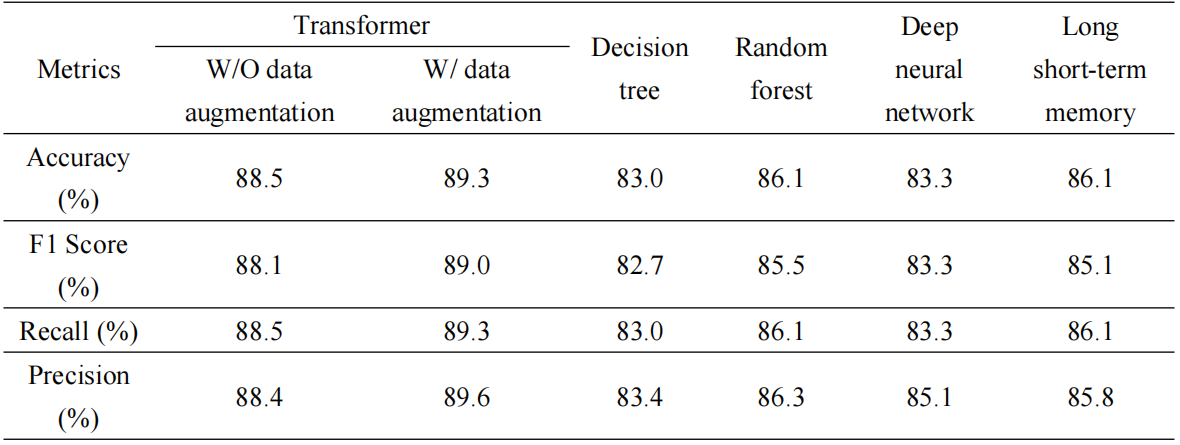}
\end{figure}

\begin{figure}[htbp]
    \centering
    \caption*{Supplementary Table 3 | Evaluating the performances of the model for predicting molar amounts of reactants using five-fold cross-validation.}
    \includegraphics[max width=0.8\textwidth]{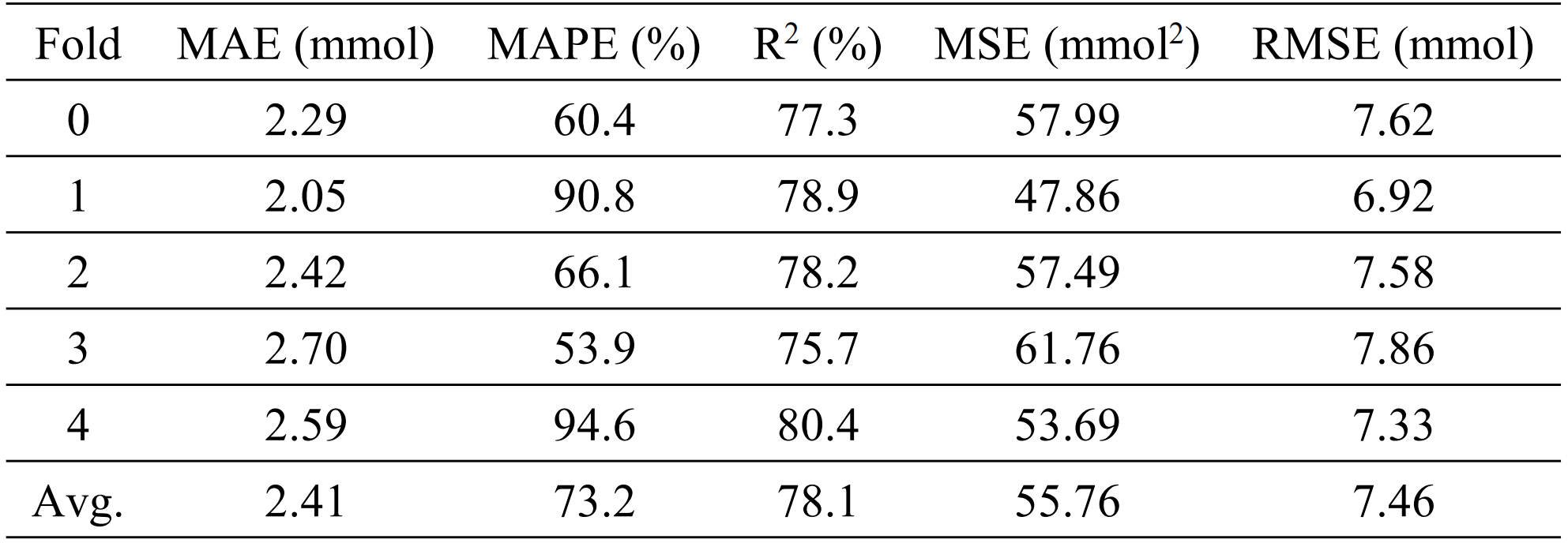}
\end{figure}

\begin{figure}[htbp]
    \centering
    \caption*{Supplementary Table 4 | Size prediction for recipes reported from 2023 to 2024. \footnote{$^a$ Equivalent circular diameters, i.e., corrected for the reported size of non-spherical nanocrystals.}}
    \includegraphics[max width=0.8\textwidth]{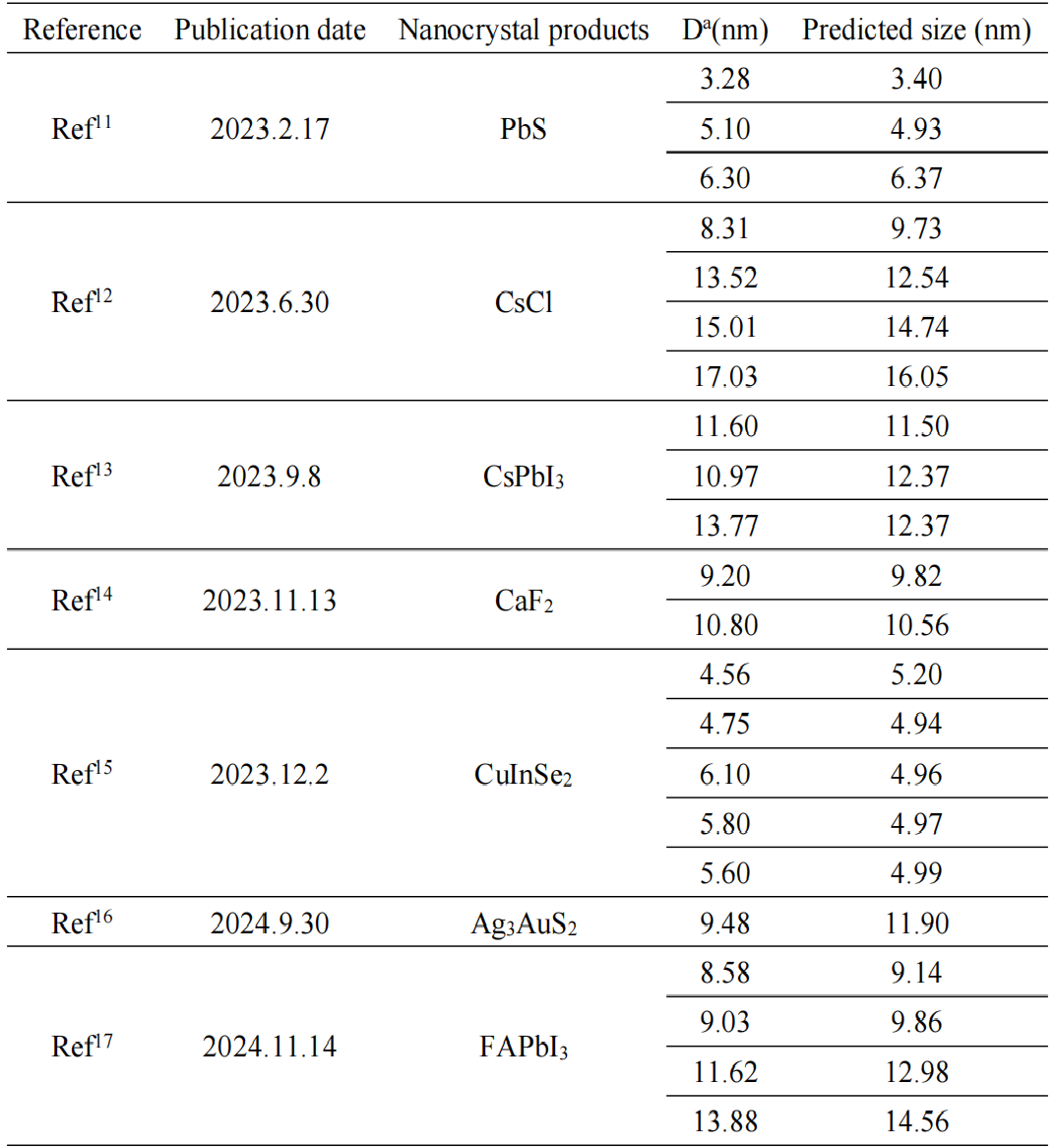}
\end{figure}

\begin{figure}[htbp]
    \centering
    \includegraphics[max width=1.0\textwidth]{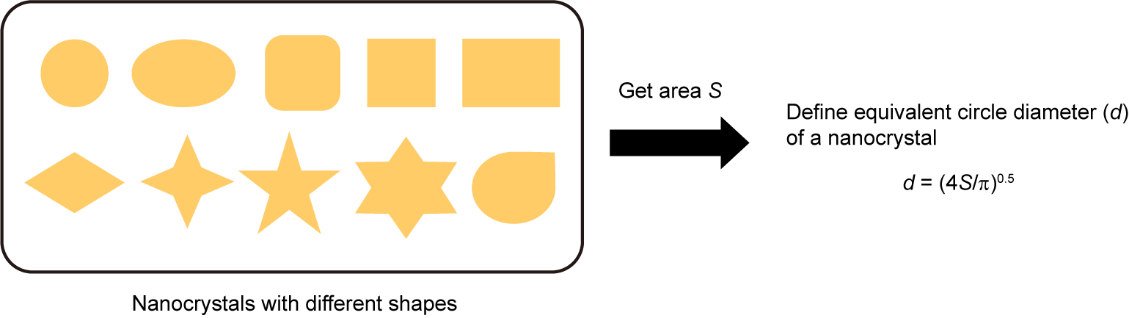}
    \caption*{Supplementary Fig. 1 | Definition of equivalent circle diameter. Based on the scale bar in a TEM image, the true area of different shapes of nanocrystals can be calculated. The equivalent circle diameter is calculated according to the equation in this figure.}
\end{figure}

\begin{figure}[htbp]
    \centering
    \includegraphics[max width=1.0\textwidth]{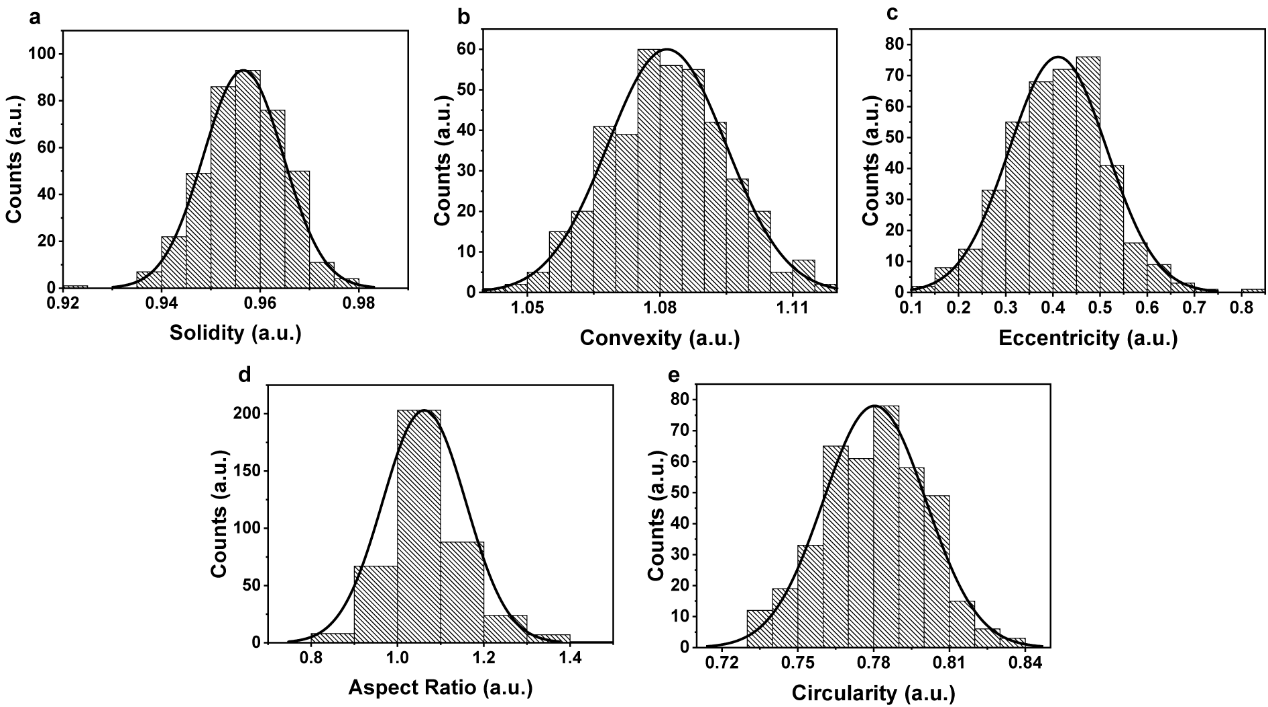}
    \caption*{Supplementary Fig. 2 | Five shape descriptors derived from the outline of segmented nanocrystals in Fig. 2b. Distributions for five shape descriptors, including solidity (a), convexity (b), eccentricity (c), aspect ratio (d), and circularity (e).}
\end{figure}

\begin{figure}[htbp]
    \centering
    \includegraphics[max width=1.0\textwidth]{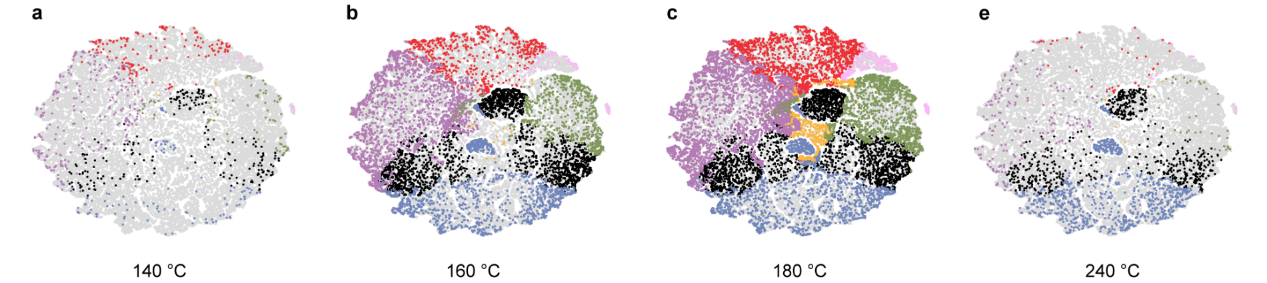}
    \caption*{Supplementary Fig. 3 | Clustering slices of 130,000 PbSe nanocrystals with different reaction temperatures in Fig. 2c. Different shapes of nanocrystals synthesized at 140 °C (a), 160 °C (b), 180 °C (c), and 240 °C (d) are distinguished by colors.}
\end{figure}

\begin{figure}[htbp]
    \centering
    \includegraphics[max width=1.0\textwidth]{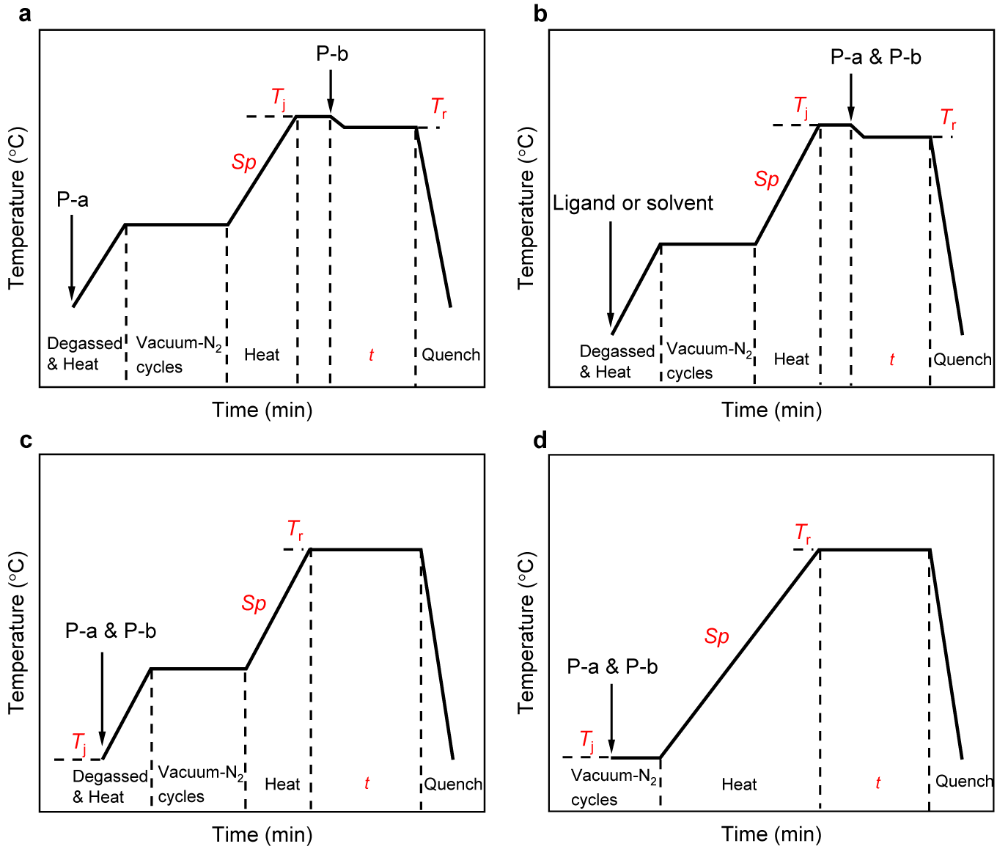}
    \caption*{Supplementary Fig. 4 | Schematics of temperature changes for hot-injection and heat-up method. Hot-injection (a, b) and heat-up method (c, d) are the most typical synthesis methods for nanocrystals. The addition order of chemicals during synthesis is ignored. Five important synthetic parameters were extracted as condition descriptors, including injection temperature ($T_j$), reaction temperature ($T_r$), heating rate ($S_p$), reaction time ($t$) and reactant molar amount ($Mol$). The abbreviations P-a and P-b in this figure represent two different precursors, a and b, respectively.}
\end{figure}

\begin{figure}[htbp]
    \centering
    \includegraphics[max width=1.0\textwidth]{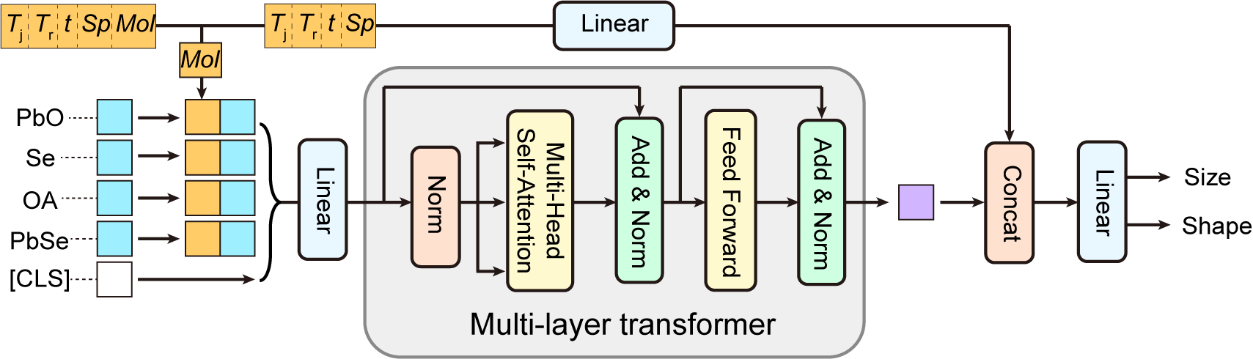}
    \caption*{Supplementary Fig. 5 | Structure of the transformer algorithm. The chemical descriptors, along with condition descriptors, were fed into the transformer model which employed a multi-layer architecture with self-attention mechanisms. The CLS token was introduced to capture global contextual information. The final predictions for size and shape were generated from a linear layer after a concat operator between CLS token and condition descriptors.}
\end{figure}

\begin{figure}[htbp]
    \centering
    \includegraphics[max width=1.0\textwidth]{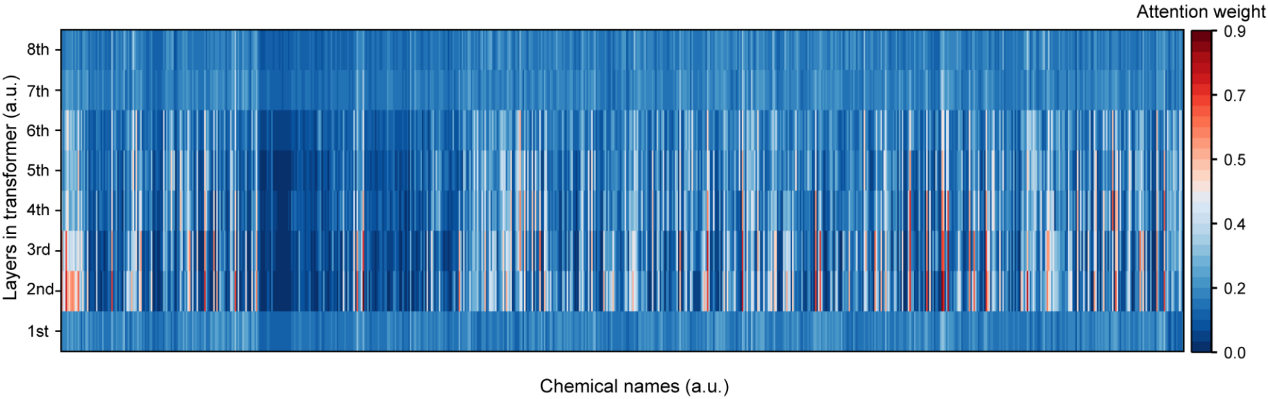}
    \caption*{Supplementary Fig. 6 | Average attention weight maps of CLS to chemicals in different transformer layers. The horizontal coordinate is the name of all chemicals in the recipe dataset (refer to source data).}
\end{figure}

\begin{figure}[htbp]
    \centering
    \includegraphics[max width=1.0\textwidth]{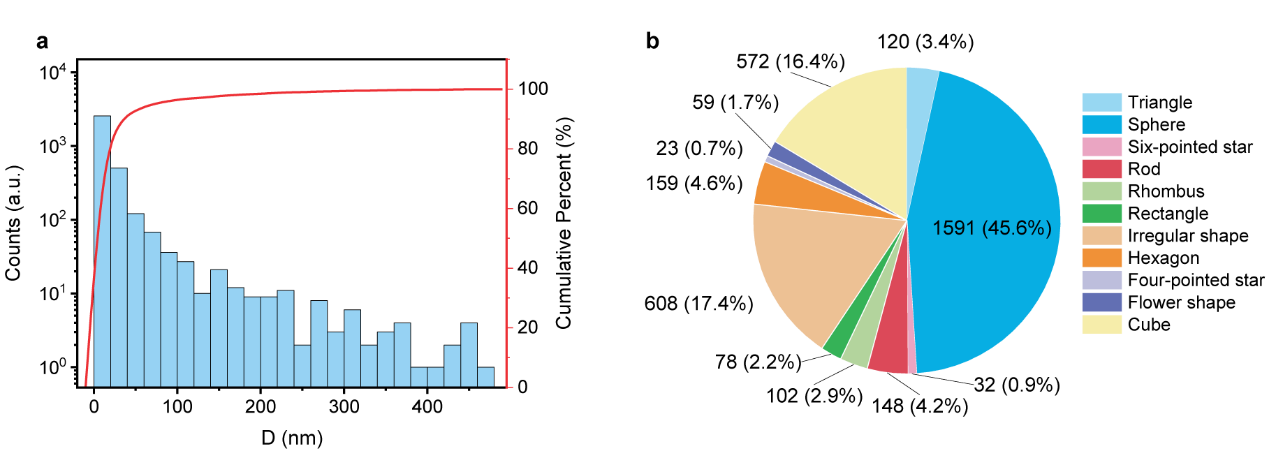}
    \caption*{Supplementary Fig. 7 | Distributions of size (a) and shape (b) labels for the synthesis model.}
\end{figure}

\begin{figure}[htbp]
    \centering
    \includegraphics[max width=1.0\textwidth]{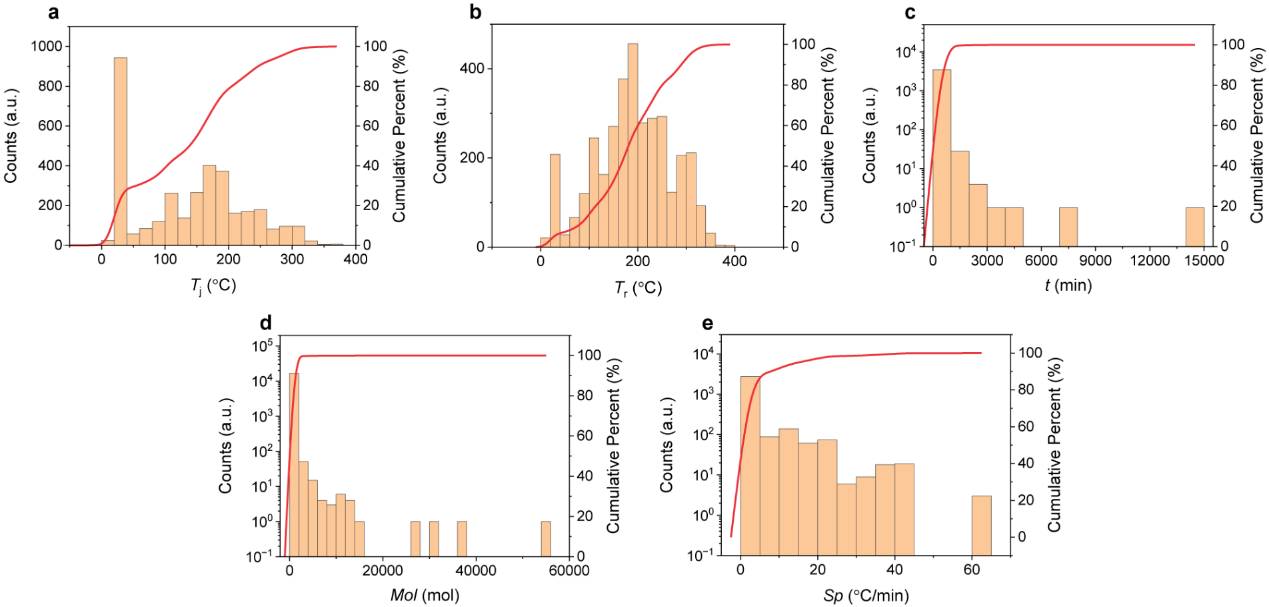}
    \caption*{Supplementary Fig. 8 | Distributions of the five synthetic parameters including injection temperature (a), reaction temperature (b), reaction time (c), reactant molar amount (d), and heating rate (e).}
\end{figure}

\newpage

\section*{Supplementary References}

\begin{enumerate}
    \item Xiao, C. et al. VISN: virus instance segmentation network for TEM images using deep attention transformer. \textit{Briefings Bioinf.} \textbf{24} (2023).
    \item Stringer, C., Wang, T., Michaelos, M. \& Pachitariu, M. Cellpose: a generalist algorithm for cellular segmentation. \textit{Nat. Methods} \textbf{18}, 100–106 (2021).
    \item Reich, C., Prangemeier, T., Françani, A. O. \& Koeppl, H. An instance segmentation dataset of yeast cells in microstructures. In \textit{2023 45th Annual International Conference of the IEEE Engineering in Medicine \& Biology Society (EMBC)} 1–4 (IEEE, 2023).
    \item Caicedo, J. C. et al. Evaluation of deep learning strategies for nucleus segmentation in fluorescence images. \textit{Cytometry Part A} \textbf{95}, 952–965 (2019).
    \item Mahbod, A. et al. NuInsSeg: A fully annotated dataset for nuclei instance segmentation in H\&E-stained histological images. \textit{Sci. Data} \textbf{11}, 295 (2024).
    \item Mahbod, A. et al. CryoNuSeg: A dataset for nuclei instance segmentation of cryosectioned H\&E-stained histological images. \textit{Comput. Biol. Med.} \textbf{132}, 104349 (2021).
    \item Depto, D. S. et al. Automatic segmentation of blood cells from microscopic slides: A comparative analysis. \textit{Tissue and Cell} \textbf{73}, 101653 (2021).
    \item Yildirim, B. \& Cole, J. M. Bayesian particle instance segmentation for electron microscopy image quantification. \textit{J. Chem. Inf. Model.} \textbf{61}, 1136–1149 (2021).
    \item Rühle, B., Krumrey, J. F. \& Hodoroaba, V.-D. Workflow towards automated segmentation of agglomerated, non-spherical particles from electron microscopy images using artificial neural networks. \textit{Sci. Rep.} \textbf{11}, 4942 (2021).
    \item Treder, K. P. et al. nNPipe: a neural network pipeline for automated analysis of morphologically diverse catalyst systems. \textit{npj Comput. Mater.} \textbf{9}, 18 (2023).
    \item Liu, Y. et al. Breaking the size limitation of directly-synthesized PbS quantum dot inks toward efficient short-wavelength infrared optoelectronic applications. \textit{Angew. Chem. Int. Ed.} \textbf{62}, e202300396 (2023).
    \item Zhuang, X. et al. Trivalent europium-doped CsCl quantum dots for MA-free perovskite solar cells with inherent bandgap through lattice strain compensation. \textit{Adv. Mater.} \textbf{35}, 2302393 (2023).
    \item Zhu, H. et al. Long-term stability in $\alpha$-CsPbI$_3$ quantum dots via a nonvacuum Ag$^+$ doping strategy. \textit{J. Phys. Chem. C} \textbf{127}, 18727–18735 (2023).
    \item Paik, T. et al. Shape-controlled synthesis and self-assembly of highly uniform upconverting calcium fluoride nanocrystals. \textit{Inorg. Chem. Front.} \textbf{11}, 278–285 (2024).
    \item Lian, W. et al. Near-infrared nanophosphors based on CuInSe$_2$ quantum dots with near-unity photoluminescence quantum yield for micro-LEDs applications. \textit{Adv. Mater.} \textbf{36}, 2311011 (2024).
    \item Han, L. et al. Ternary Ag$_3$AuS$_2$ nanocrystals for thin-film solar cells. \textit{Inorg. Chem.} \textbf{63}, 19382–19389 (2024).
    \item Sun, X. et al. Diffusion-mediated synthesis of high-quality organic–inorganic hybrid perovskite nanocrystals. \textit{Nat. Synth.} (2024).
    \item Zheng, W. et al. Ultra stable X-Ray imaging through a mutually reinforcing strategy between perovskite nanocrystal-polymethyltrifluoropropylsiloxane. \textit{Adv. Funct. Mater.}, 2418944 (2024).
\end{enumerate}

\end{document}